\documentclass[aps,prd,nofootinbib,superscriptaddress,twocolumn]{revtex4}

\usepackage{slashed}
\usepackage{hyperref}
\usepackage{amsmath,amsfonts,amssymb}
\usepackage{booktabs}
\usepackage{comment}
\usepackage{siunitx}
\usepackage{xcolor}
\usepackage{verbatim}
\usepackage{slashed}
\usepackage[english]{babel}
\usepackage{caption}
\usepackage{subcaption}
\usepackage{multirow}

\usepackage{graphicx}

\newcommand{\be}{\begin{equation}}
\newcommand{\ee}{\end{equation}}
\newcommand{\bea}{\begin{equation}\begin{aligned}}
\newcommand{\eea}{\end{aligned}\end{equation}}
\newcommand{\nn}{\nonumber}

\newcommand{\sdfrac}[2]{\mbox{\small$\displaystyle\frac{#1}{#2}$}}

\newcommand{\gsim}{\lower.7ex\hbox{$\;\stackrel{\textstyle>}{\sim}\;$}}
\newcommand{\lsim}{\lower.7ex\hbox{$\;\stackrel{\textstyle<}{\sim}\;$}}

\begin{document}

\title{Beyond the Inert Doublet:\\ imprints of Scotogenic Yukawa interactions at FCC-ee}

\author{Carlo Marzo}
\affiliation{Laboratory for High Energy and Computational Physics, NICPB, R\"{a}vala 10, Tallinn 10143, Estonia}
\author{Aurora Melis}
\affiliation{Istituto Nazionale di Fisica Nucleare, Sezione di Roma Tre, Via della Vasca Navale 84, I-00146 Rome, Italy}

\date{\today}

\begin{abstract}
It is tempting to interpret the minuscule scale of neutrino masses as a symptom of its radiative origin. 
In light of the notable leap in precision expected at the Future Circular Collider, we explore areas of the parameter space that can simultaneously support the detectable Higgs-strahlung signal with parallel ones from forthcoming measurements in low-energy observables. 
We pinpoint the role that the extra fermions have in shaping a signal distinct from the pure Inert Doublet one. The details of the full one-loop computation and on-shell renormalization are presented. Both normal and inverted hierarchies for the radiatively generated neutrino masses and angles are investigated.
\end{abstract}

\maketitle


\section{Introduction}
\label{sec:Intro}

The exploration of the fundamental interactions through the lenses of the Large Hadron Collider (LHC) has cemented the Standard Model (SM) as the renormalizable theory of the Electro-Weak (EW) scale. The profiling of the scalar sector, crowned by the Higgs discovery, has generated an abundance of challenging constraints for alternatives and extensions of the SM. Nevertheless, the incomplete status of the SM is still sustained by the pressing evidence of neutrino oscillations, dark matter or the strong-CP problem, to cite a few hints for New Physics (NP). 
A further experimental effort, in synergy with a corresponding theoretical one, is therefore of paramount importance to seek clarity towards such puzzling observations.

The discovery-first motivation behind the building of hadron colliders, reaching well into the TeV energy scale, naturally contemplates a complementary attempt towards high-precision measurements.
In light of this the experimental community has devised diverse novel realizations for a future electron-positron collider, which would exploit the clean leptonic initial state to provide precise profiling of particle dynamics \cite{ILC:2013jhg,Asner:2013psa,CEPCStudyGroup:2018ghi,An:2018dwb,Fan:2014vta,FCC:2018evy,Blondel:2018mad}. 
The Future Circular Collider is one promising realization of this shift towards precision.
It is designed, in one of its available setups (FCC-ee), to collide electron-positron pairs at centre-of-mass energy from around the $Z$ pole to the range 240 and 365 GeV, where Higgs production will take place via Higgs-strahlung $e^+ e^- \rightarrow Z h$ and $WW$ boson fusion $e^+ e^- \rightarrow h \nu_e \bar{\nu}_e$.
The FCC-ee is expected to collect over a million of $Z h$ events and one order of magnitude less for the single Higgs event produced by WW fusion. Interestingly, in $Z h$ final states, the Higgs-production cross-section will be determined independently of the modelling of the Higgs decay.
With perfect efficiency, this would serve the theory community with a statistical precision of per-mil level over the $Zh$ production cross-section. 
A slightly more prudent precision ($0.5\%$) is expected when simulating the collision with realistic detectors \cite{Azzurri:2021nmy}. 

Even in such conservative scenarios, the FCC-ee will be sensitive to deviations from the SM predictions whose size are comparable to that of radiative corrections, hence, to the scale of NP (virtual) particles. 
To translate the cutting-edge experimental design into discovery potential, the theory community must match the statistical precision with, at least, a comparably precise theoretical determination.
The profiling of the dominant next-to-leading (NLO) \cite{Fleischer:1982af,Kniehl:1991hk,Denner:1992bc} and next-to-next-to-leading (NNLO) \cite{Freitas:2022hyp} EW radiative signature in $e^+ e^-\rightarrow Z h$, which in turn demands a rigorous inclusion of QED corrections, enables to pinpoint the presence of NP in multiple observables, ultimately connected to the differential production cross-section (see for instance \cite{Craig:2014una,Gounaris:2014tha}). 
Once full control over the SM background is established, it is therefore natural to look for proper signatures coming from minimal, viable extensions of the SM. In such direction, the recent analysis of \cite{Abouabid:2020eik} has explored the radiative imprints over $e^+ e^-\rightarrow Z h$ generated by an extra, inert, $SU(2)_L$ scalar doublet. Interestingly, such a minimal scenario, which is also a commonly shared feature of more involved modifications of the SM scalar sector, can leave a signature that can be detected at FCC-ee while still surviving the tight bounds from LHC.  

In this work, we continue this investigation by assessing the impact of a minimal modification of the Yukawa sector. We do this considering the role of extra fermions, neutral under SM gauge transformations, and connected to the SM particles via an inert doublet. We consider whether the presence of extra Yukawa couplings can modify the expected yield in $e^+ e^-\rightarrow Z h$ generated by the presence of the inert scalar, and therefore the possibility to discriminate (obviously supported by a complementary set of observations, as in any realistic discovery scenario) between the minimal IDM and its extensions. While some of our computations and conclusions are general enough to comprise the effects of the extra Yukawa couplings, we enrich the investigation by exploring a connection with neutrino phenomenology and related flavour observables. We consider the Scotogenic model \cite{Ma:2006km}, where the texture of the new Yukawa interactions, as well as the coupling and masses of the scalar sector, cooperate to radiatively generate the operators fitting neutrino phenomenology. 
Such connection is realized for areas of the parameter space that are naturally subjected to the tight constraints of leptonic flavor violation (LFV), as well as other bounds, which we extensively confront in a synergic attempt to enhance their $e^+ e^-\rightarrow Z h$ signature. 
We limit our analysis to real NP parameters, postponing to future scrutiny the inclusion of CP-violating effects.

\section{The Model} 
\label{sec:Model}
In the Scotogenic model, the SM particle content is augmented by three singlet Right Handed Neutrinos (RHN) $N_{k=1,2,3}$ and one $SU(2)_L$ scalar doublet, $H_2$, in addition to the SM-Higgs doublet, therefore denoted by $H_1$. 
The new fields are odd under an auxiliary $\mathbb{Z}_2$ discrete symmetry while all the SM fields are even. 
The Lagrangian of the model coincides with the one of the Inert Doublet Model (IDM) supplemented with the interactions of the RHN sector:
\bea
\label{eq:LagScoto}
\mathcal{L}^{\rm Scoto} = \mathcal{L}^{\rm IDM} + \mathcal{L}^{\rm RHN} ,
\eea
where
\begin{align}
\label{eq:LagIDM}
\mathcal{L}^{\rm IDM} = \mathcal{L}^{\rm SM} &+ |D_\mu H_2|^2 - \mu^2_2 |H_2|^2 - \lambda_2 |H_2|^4 \nn \\
&-\lambda_3 |H_1|^2 |H_2|^2 -\lambda_4 |H^\dagger_1 H_2|^2 \\
&-\frac{\lambda_5}{2} \left[ (H_1^\dagger H_2)^2+ {\rm h.c.}\right] ,\nn
\end{align}
\bea
\label{eq:LagRHN}
\mathcal{L}^{\rm RHN} = \overline{N}_k \slashed{\partial} N_k - y^N_{ik} \overline{L}_i \widetilde{H}_2 N_k -\frac{m_{N_k}}{2}\overline{N^c_k} N_k + {\rm h.c.} .
\eea
Here $L_i=(\nu_i,l_i)^T_L$ is the lepton doublet and $\widetilde{H}_2=i\sigma_2 H^*_2$ with $\sigma_2$ the second Pauli matrix. A large portion of the parameter space is compatible with the inert minimum where only the SM doublet acquires the vacuum expectation value (vev) $v_{h}/\sqrt{2}$. Expanding around this minima, the two scalar doublets can be parametrized as
\bea
H_1 = 
\begin{pmatrix} 
G^+ \\ 
\sdfrac{v_{h} + h + i G^0}{\sqrt{2}}
\end{pmatrix}\,,\quad
H_2 = 
\begin{pmatrix} 
H^+ \\ 
\sdfrac{ H^0 + i A^0}{\sqrt{2}}
\end{pmatrix}\, ,
\eea
where $G^0, G^\pm$ are the Nambu-Goldstone bosons associated with the spontaneous symmetry breaking of the electroweak symmetry, which causes the SM-Higgs $h$ to appear in the scalar spectrum along with four other scalar particles: one neutral scalar (CP-even) $H^0$, one pseudoscalar (CP-odd) $A^0$ and a couple of charged scalars $H^\pm$. 
At tree level, we express the masses of the new scalar particles in terms of the average neutral scalar mass $m^2_0 \equiv \tfrac{1}{2} (m^2_{H^0} + m^2_{A^0}) =\mu^2_2 + (\lambda_3 + \lambda_4) \tfrac{v^2_h}{2}$ as
 \bea
 \label{eq:scalarmasses}
 m^2_{H^\pm} &= m^2_0 -\lambda_4 \tfrac{v^2_h}{2}\,, \\
 m^2_{H^0} &= m^2_0 +\lambda_5 \tfrac{v^2_h}{2}\,, \\
 m^2_{A^0} &= m^2_0 -\lambda_5 \tfrac{v^2_h}{2} \,. \\
 \eea
 In the case $|\lambda_5|\ll |\lambda_4|$, typically preferred in the Scotogenic model as explained in Section~\ref{sec:NeuSection}, the value of $\lambda_4$ controls the splitting between the charged and neutral components, while $\lambda_5$ governs the splitting between the CP-even and -odd neutral scalars.
 
A noteworthy list of theoretical and experimental bounds severely restricts the parameter space of the Scotogenic model. We go over those that are most important to our analysis.

 \subsection{Constraints from vacuum stability and perturbativity}
 \label{sec:VacuumStabilityandPerturbativity}
 The consistency of the perturbative calculations showed in this paper requires at least $\lambda_i < 8 \pi $ to guarantee the perturbativity of the quartic couplings. Moreover, the condition that the scalar potential must be bounded from below translates into the following restrictions 
 \bea
 \label{eq:Llimits}
 \lambda_{2}>0, \quad \lambda_3\,,\, \lambda_3 + \lambda_4 -|\lambda_5|> -2\sqrt{\lambda_1 \lambda_2},
 \eea
 where $\lambda_1$ is the SM quartic coupling whose value is fixed by the SM-Higgs mass and its vacuum expectation value: $\lambda_1=m^2_{h}/2 v^2_h\simeq0.132$.\\
The requirement of perturbative unitarity of all the scattering processes involving the inert scalars leads to limiting the eigenvalues $e^\pm_i$ of the full scattering matrix $|e^\pm_i|< 8\pi$:
 \begin{eqnarray}
     &&e^\pm_1 =\lambda_3\pm\lambda_{4,5}\,,\quad e_2 = \lambda_3 +2 \lambda_4 \pm 3 \lambda_5 \,, \nn\\
     &&e^\pm_4 = -(\lambda_1 +\lambda_2) \pm \sqrt{(\lambda_1-\lambda_2)^2+\lambda_{4,5}^2 }\,,\\
     &&e^\pm_5 = -3(\lambda_1 +\lambda_2) \pm \sqrt{9(\lambda_1-\lambda_2)^2+(2\lambda_3+\lambda_4)^2 }.\nn
\end{eqnarray}
 An upper limit over $\lambda_2$ is obtained from $e^-_5$ when $\lambda_3=\lambda_4=0$, that is $\lambda_2 < 4\pi/3$, which plugged into Eq.~\ref{eq:Llimits} gives $\lambda_3>-\sqrt{8\pi/3} (m_h/v_h)\simeq -1.49$.

\subsection{Constraints from Neutrino Phenomenology} 
\label{sec:NeuSection}
\begin{figure}[t]
\centering
\includegraphics[scale=0.14]{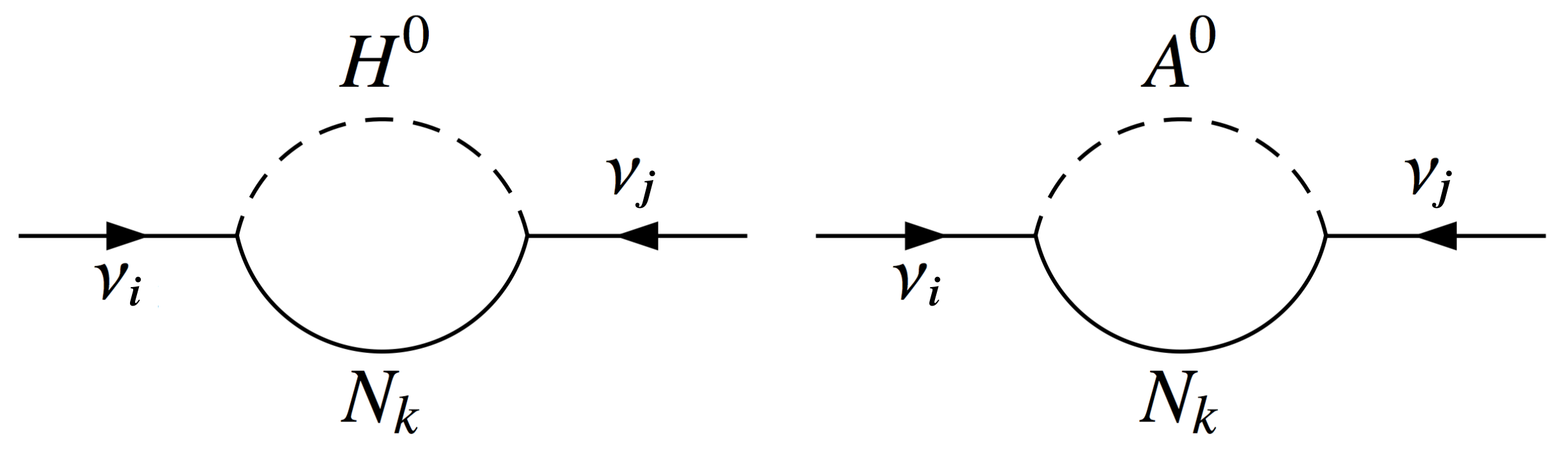}
\caption{One-loop Feynman diagrams for the neutrino mass generation. }
\label{fig:numass}
\end{figure}
In the Lagrangian of Eq.(\ref{eq:LagScoto}), a $\mathbb{Z}_2$ symmetry precludes the possibility to write the term $\overline{L}_i H_1 N_k$, so that at tree-level neutrinos are massless. Their mass is radiatively generated at the one-loop level through the diagrams in Fig. \ref{fig:numass}, left from the spontaneous symmetry breaking of the dimension-5 operator  $\overline{L}_i \cdot \tilde{H}_1 H^T_1 \cdot L_j^c $ \cite{Weinberg:1979sa}. This gives
\bea
\label{eq:mnu}
&(m_\nu)_{ij} = (Y_N \Lambda Y^T_N)_{ij}= \sum^3_{k=1} y^N_{ik}y^N_{jk} \Lambda_k \\
\Lambda_{k} = &\frac{m_{N_k}}{16\pi^2} \bigg(B_0[0,m^2_{H^0},m^2_{N_k}]-B_0[0,m^2_{A^0},m^2_{N_k}]\bigg),
\eea
where the function $B_0$ is 
\bea
B_0[0,m^2,m^2_{N}] &= \sdfrac{1}{\epsilon_{\rm UV}}+1 - \sdfrac{m^2_{0}}{m^2-m^2_{N}} \ln{\sdfrac{m^2}{m^2_{N}}} +\ln{\sdfrac{\mu^2}{m^2_{N}}}.
\eea
The crucial point of the Scotogenic construction is that, since the scalar and pseudoscalar diagrams give equals but opposite in sign contributions, the size of the neutrino masses is controlled by the mass splitting $m^2_{H_0}-m^2_{A_0}=\lambda_5 v^2_h $. This is understood by expanding $\Lambda_k$ in small $\lambda_5$:
\bea \label{eq:mnuExp}
\Lambda_{k}=\frac{\lambda_5 v^2_h}{16\pi^2} \frac{ m_{N_k}}{m^2_0-m^2_{N_k}}\left[1-\sdfrac{m^2_{N_k}}{m^2_{0}-m^2_{N_k}}\ln{\sdfrac{m^2_{0}}{m^2_{N_k}}}\right] +\mathcal{O}(\lambda_5^2)\,.
\eea
In the approximation of $m_0 \simeq m_{N_k}$ the above loop function tends to $1/2 m_{N_k}$ giving 
\bea
(m_{\nu})_{ij}\simeq\frac{\lambda_5 v^2_h}{32 \pi^2} \sum^3_{k=1} \frac{y^N_{ik}y^N_{jk}}{m_{N_k}}.
\eea
From this, it is clear that the size of the typical see-saw scale is reduced by a factor of $\lambda_5 v^2_h/32\pi^2$, with a proper choice of $\lambda_5$ this scale can be as low as to be accessible at colliders. This is what makes the Scotogenic model highly attractive. In our numerical analysis, we assume $\lambda_5 \simeq \{10^{-12}-10^{-9}\}$ so that the neutrino scale is naturally placed at the eV scale with $\mathcal{O}(1)$ Yukawa couplings and GeV RHN masses. 
\begin{table}[t]
\centering
     {\renewcommand{\arraystretch}{1.5} 
     \resizebox{0.8\columnwidth}{!}{
        \begin{tabular}{|c c c|}
        \hline
        \bf Neutrino observable & \bf NH & \bf IH\\ 
        \hline
        $\Delta m_{21}^{2}$ [$10^{-5}$eV$^{2}$]  & $7.42^{+0.21}_{-0.20}$ & $
        7.42^{+0.21}_{-0.20}$  \\ 
        $\Delta m_{3l}^{2}$ [$10^{-3}$eV$^{2}$]  & $+2.515_{-0.028}^{+0.028}$ & $
        -2.498_{-0.029}^{+0.028}$  \\ 
        $\sin^2\theta_{12}$  & $0.304^{+0.013}_{-0.012}$ & $0.304^{+0.012}_{-0.012}$  \\ 
        $\sin^2\theta_{13}$  & $0.02220^{+0.00068}_{-0.00062}$ & $0.02238^{+0.00064}_{-0.00062}$
        \\
        $\sin^2\theta_{23}$  & $0.573^{+0.018}_{-0.023}$ & $0.578^{+0.017}_{-0.021}$ \\ 
        $\delta_{\rm CP}[^{\circ }]$  & $194^{+52}_{-25}$ & $287^{+27}_{-32}$  \\ 
        \hline
        \end{tabular}
         }
     }
\caption{ \label{tab:neutrinoData} \small Global best-fit values for the neutrino mass squared splittings, mixing angles, and CP-violating phase. The
experimental values are taken from Ref.~\cite{Esteban:2020cvm}}
\end{table}

The resulting neutrino mass matrix has to be compatible with the neutrino oscillation data in Table~\ref{tab:neutrinoData}. In particular, it has to be diagonalized by the Pontecorvo-Maki-Nakawaga-Sakata (PMNS) mixing matrix $U^\nu$: $\widehat{m}_\nu \equiv {\rm diag}(m_{1},m_{2},m_{3})= U^{\nu T} m_\nu U^\nu$. 
Enforcing this, in the basis where the RHN mass matrix is diagonal, the most general Yukawa matrix can be written in terms of the Casas-Ibarra parametrization \cite{Casas:2001sr}
\bea
\label{eq:CasasIbarra}
Y_N=U^{\nu *} \sqrt{\widehat{m}_\nu}R\sqrt{\Lambda_k}^{-1}\,,
\eea
where $R$ is an arbitrary complex-orthogonal matrix ($R^TR = \mathbb{I}$). Each entry of the Yukawa matrix reads 
\begin{eqnarray} \label{eq:CasasIbarraParam}
    y^N_{ik} = \frac{\sqrt{ m_{_1}}\,U^{\nu*}_{i1}R_{1k}  + \sqrt{m_{2}}\,U^{\nu*}_{i2} R_{2k} + \sqrt{m_{3}} \,U^{\nu*}_{i3} R_{3k} }{\sqrt{\Lambda_k}}\,.
\end{eqnarray}
 The current neutrino global fits in Ref.~\cite{Esteban:2020cvm} are compatible with either a Normal Hierarchy (NH) scenario for neutrino masses or an Inverted Hierarchy (IH) scenario, with a preference for the first alternative. Specifically, in the NH-scenario $m_{\rm min}\equiv m_{1}$ with $m_{1}<m_{2}<m_{3}$ 
\begin{eqnarray}
    m_{_2} = \sqrt{\Delta m^2_{21}+m^2_{1} } \,, \quad m_{3} = \sqrt{\Delta m^2_{31}+m^2_{1} },
\end{eqnarray}
while in the IH-scenario $m_{\rm min}\equiv m_{3}$ with $m_{3}<m_{1}\simeq m_{2}$
\begin{eqnarray}
    m_{1} = \sqrt{\Delta m^2_{32}-\Delta m^2_{21}+m^2_{3}} \,, \quad m_{2} = \sqrt{\Delta m^2_{32}+m^2_{3} }.
\end{eqnarray}
Assuming $m_{\rm min}\simeq 0$ and the entries of the $R$ matrix of the same order we can deduce that
\bea
\frac{(y^{N}_{ik})^{\rm IH}}{(y^{N}_{ik})^{\rm NH}} \simeq \frac{U^{\nu*}_{i1}R_{1k}+U^{\nu*}_{i2}R_{2k}}{U^{\nu*}_{i3}R_{3k}}.
\eea

\begin{table}[t]
	\centering
	{\renewcommand{\arraystretch}{1.5}
	\resizebox{1\columnwidth}{!}{
	\begin{tabular}{|c c c|}
    \hline
    \bf LFV Process & \bf Current Limit & \bf Future Limit \\[2.pt]
    \hline
        BR($\mu\rightarrow e \gamma$) & $4.2\times10^{-13}$ (\texttt{MEG at PSI}) & $6\times10^{-14}$ \texttt{(MEG\,II}) \\
        BR$(\tau\rightarrow e\gamma)$  & $3.3\times 10^{-8}$ (\texttt{BaBar}) & \qquad $5\times10^{-9}$ (\texttt{Belle\,II}) \\
        BR$(\tau\rightarrow\mu\gamma)$ & $4.4\times 10^{-8}$ (\texttt{BaBar}) & \qquad $10^{-9}$ (\texttt{Belle\,II}) \\
        BR($\mu\rightarrow 3 e$)     & $1.0\times10^{-12}$ (\texttt{SINDRUM}) & ~~\quad $10^{-16}$ (\texttt{Mu3e}) \\
        BR$(\tau \rightarrow 3 e)$  & $2.7\times10^{-8}$ (\texttt{Belle}) & \qquad $5\times10^{-10}$ (\texttt{Belle\,II}) \\
        BR$(\tau\rightarrow 3\mu)$   & $2.1\times10^{-8}$ (\texttt{Belle}) & \qquad $5\times10^{-10}$ (\texttt{Belle\,II}) \\
        \hline
        BR$(Z\rightarrow \mu e)$   & $7.5 \times 10^{-7}$ (\texttt{LHC ATLAS}) & $10^{-10}-10^{-8}$ (\texttt{FCC-ee}) \\
        BR$(Z\rightarrow \tau e)$   & $9.8 \times 10^{-6}$ (\texttt{LEP OPAL}) & $10^{-9}$ (\texttt{FCC-ee}) \\
         BR$(Z\rightarrow \tau \mu)$   & $1.2 \times 10^{-5}$ (\texttt{LHC DELPHI}) &  $10^{-9}$(\texttt{FCC-ee}) \\

        BR$(h\rightarrow \mu e)$   & $6.1 \times 10^{-5}$ (\texttt{LHC CMS}) &  $-$ \\
        BR$(h\rightarrow \tau e)$   & $4.7 \times 10^{-3}$ (\texttt{LHC CMS}) &  $-$ \\
         BR$(h\rightarrow \tau \mu)$   & $2.5 \times10^{-3}$ (\texttt{LHC CMS}) &  $-$ \\        
    \hline
    \end{tabular}}}
\caption{\label{tab:LFV}\small Current and future expected limits on the LFV processes considered in this paper.}
\end{table}

Only the first line ($i=1$) of the RHN Yukawa matrix enters our targeted process, $e^+e^-\rightarrow Zh$. Being $|U^\nu_{11}|\sim 0.82$, $|U^\nu_{12}|\sim 0.55$ $|U^\nu_{13}|\sim 0.15$, from the above reasoning we can typically expect larger Yukawa entries in the IH scenario when compared to the NH one. 

In order to replicate the observed neutrino mixing, the new RHN Yukawa must exhibit a non-trivial structure with off-diagonal entries of the same size as the diagonal ones. This situation is a risky source of Lepton Flavor Violation (LFV).

\subsection{Constraints from Lepton Flavor Violation}
\label{sec:LFV}

LFV searches will enter a new age marked by significant experimental efforts on numerous fronts. Table~\ref{tab:LFV} summarizes the current and future expected limits on the LFV processes considered in this paper. Presently, the most severe constraint comes from the non-detection of $\mu\rightarrow e \gamma$ at the \texttt{MEG} experiment. In the future, the precision of these observables will increase by at least one order of magnitude, in particular, we stress the extraordinary four orders of magnitude improvement expected for the $\mu\rightarrow 3 e$ process at \texttt{Mu3e}.

In this work, we employ the results of Ref.~\cite{Toma:2013zsa, Hundi:2022iva} where the one-loop analytic expressions for BR$(l_i\rightarrow l_j\gamma)$, BR$(l_i\rightarrow 3l_j)$ and BR$(Z(h)\rightarrow l_jl_j)$ have been computed within our same Scotogenic framework. 
We find that BR$(\mu\rightarrow e\gamma)$ poses the most severe constraint on our results. However, the $R$ matrix in the Casas-Ibarra parametrization of Eq.~\ref{eq:CasasIbarra} provides additional freedom to fix the entries of the RHN Yukawa. We use this freedom to elude the $\mu\rightarrow e\gamma$ constraint by fixing its value within a window fully testable by the forthcoming \texttt{MEG II} experiment.

\begin{table}[t]
\centering
     {\renewcommand{\arraystretch}{1.5} 
     \setlength{\tabcolsep}{19pt}
     \resizebox{0.8\columnwidth}{!}{
        \begin{tabular}{|c c|}
        \hline
        \bf Process & \bf Value  \\ 
        \hline
        $\Gamma(Z\rightarrow e^+e^-)$ [MeV] & $83.91 \pm 0.12$  \\
        $\Gamma(Z\rightarrow \mu^+\mu^-)$ [MeV] & $83.992 \pm 0.18$   \\
        $\Gamma(Z\rightarrow \tau^+\tau^-)$ [MeV] & $84.08 \pm 0.22$   \\
        $\Gamma(Z\rightarrow inv.)$ [MeV] & $499.0 \pm 1.5$  \\
        $\Gamma(W^+\rightarrow e^+\overline{\nu}_e)$ [MeV] & $223.3 \pm 5.6$  \\
        $\Gamma(W^+\rightarrow \mu^+\overline{\nu}_\mu)$ [MeV] & $221.6 \pm 5.4$   \\
        $\Gamma(W^+\rightarrow \tau^+\overline{\nu}_\tau)$ [MeV] & $237.3 \pm 6.5$  \\
        $\cfrac{{\rm BR}(h\rightarrow \gamma\gamma)}{{\rm BR}(h\rightarrow \gamma\gamma)^{\rm SM}}$& $1.10 \pm 0.07 $ \\
         BR$(h\rightarrow inv.)$ & $< 19$ \%  \\
        \hline
        \end{tabular}
         }
     }
\caption{ \small Particle Data Group best-fit values for the processes considered in this paper. }
\label{tab:contsraints} 
\end{table}

\subsection{Constraints from LEP and LHC}
\label{sec:LEPandLHC}
Crucial constraints on the parameters of the scalar sector of the Scotogenic model derive from the lack of NP signals from searches at LEP and LHC. Many of them coincide with those typically considered in the sole IDM model and are extensively discussed in Ref.~\cite{Belyaev:2016lok}.
In particular, the precise measurements of the total decay widths of the $Z$ and $W$ bosons at LEP forbids the decays $Z\rightarrow H^+ H^-$, $Z\rightarrow H^0 A^0$ and $W^\pm\rightarrow H^{\pm} H^0 (A^0)$. This translates into the following conditions over the new scalar masses
\begin{eqnarray}
   & 2 m_{H^{\pm}}>m_Z \,, \quad  m_{H^0}+m_{A^0}>m_Z\,,\nn\\
    & m_{H^\pm}+m_{H^0,A^0}>m_W\,.
\end{eqnarray}
Moreover, direct production searches of charginos at LEP II can be reinterpreted into a search for the new scalar $H^{\pm}$-field, where the translation from fermions to scalars weakens the bound. The conversion to a mass bound on $H^{\pm}$ gives 
\begin{eqnarray}
    & m_{H^{\pm}}> \{70-90\}\,\text{GeV} \text{\cite{Pierce:2007ut}}
\end{eqnarray}
On the contrary, the limits coming from the reinterpretation of the neutralino searches in Ref.~\cite{Lundstrom:2008ai} do not apply to nearly degenerate neutral scalars of mass splittings $|m_{H^0}-m_{A^0}| < 8\,\text{GeV}$ that is the scenario considered in this work.

Other relevant constraints come from the Higgs data at the LHC. In fact, if $m_{H^0}, m_{A^0} < m_h/2$, the invisible decays $h\rightarrow H^{0} H^{0}$ and $h\rightarrow A^{0} A^{0}$ are open. In this case, ATLAS and CMS searches set an upper bound for the branching ratio of $h\rightarrow inv.$, as given in Table~\ref{tab:contsraints}. This, in turn, reduces to a strong limit on the value of the coupling $\lambda_3+\lambda_4\pm \lambda_5$. We exclude this situation by taking $m_{H^0,A^0}>m_h/2$.

\begin{figure}[t]
\centering
\includegraphics[scale=0.20]{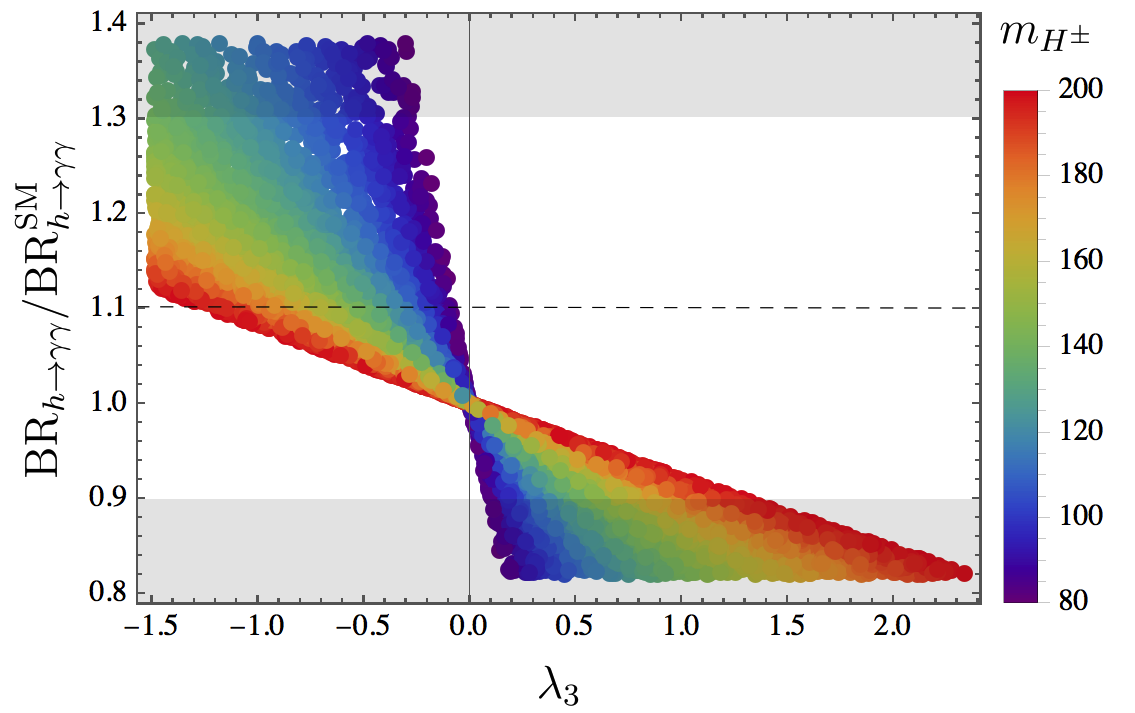}
\caption{\small The dependence of $h\rightarrow \gamma\gamma$ from the quartic coupling $\lambda_3$. The gray bands are the excluded regions imposing the $2\sigma$ experimental constraint. }
\label{fig:Rgg_vs_L3}
\end{figure}

Beyond the tree level, the di-photon decay channel $h\rightarrow\gamma\gamma$ receives radiative contributions from loops involving the charged inert scalar $H^{\pm}$. Ultimately, this results in a constraint on the value of $\lambda_3$. This is shown in Figure~\ref{fig:Rgg_vs_L3} where, for $ m_{H^\pm}\simeq \{80-200\}$~GeV, we have that  $\lambda_3< 1.4$.

Electroweak precision tests (EWPT) represent a strong constraint for the dynamics of heavy new particles in NP models. In this work, we move beyond the standard implementations of the EWPT limits \cite{Maksymyk:1993zm}. This is required by the inclusion in our study of unsuppressed, but perturbative, NP couplings to light fermions and a NP scale comparable with the mass of $Z$ and $W$ gauge bosons. Thus, in order to make contact with experiments, we explicitly compute and address the NP effects over the lepton diagonal decays $Z\rightarrow l^+l^-$, $Z\rightarrow \nu_l\nu_l$ and $W^\pm\rightarrow l^\pm\nu_l$,  which receive NP radiative corrections from the diagrams in 
\begin{figure}[t]
\centering
\includegraphics[scale=0.18]{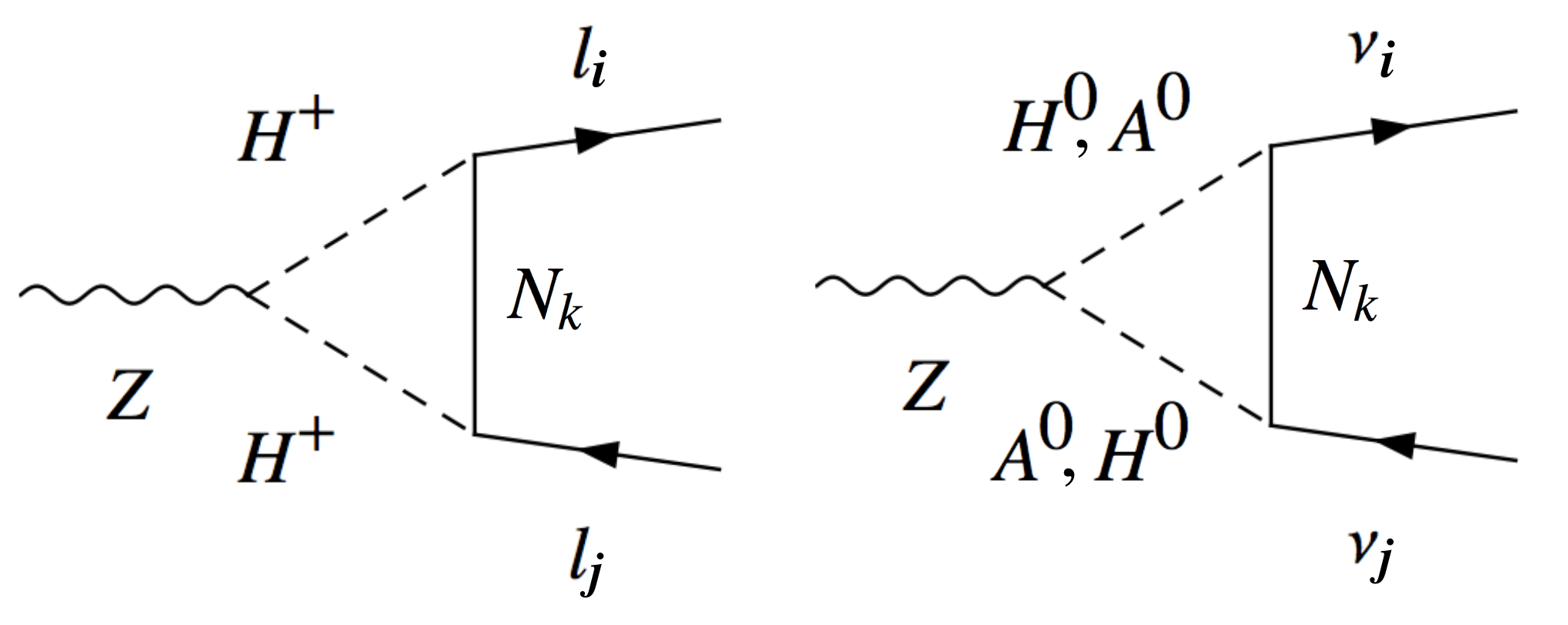}
\caption{\small One-loop corrections to $Z\rightarrow l_i l_j$ and  $Z\rightarrow \nu_i \nu_j$ in the Scotogenic model.}
\label{fig:Zll_Znunu}
\end{figure}
\begin{figure}[t]
\centering
\includegraphics[scale=0.24]{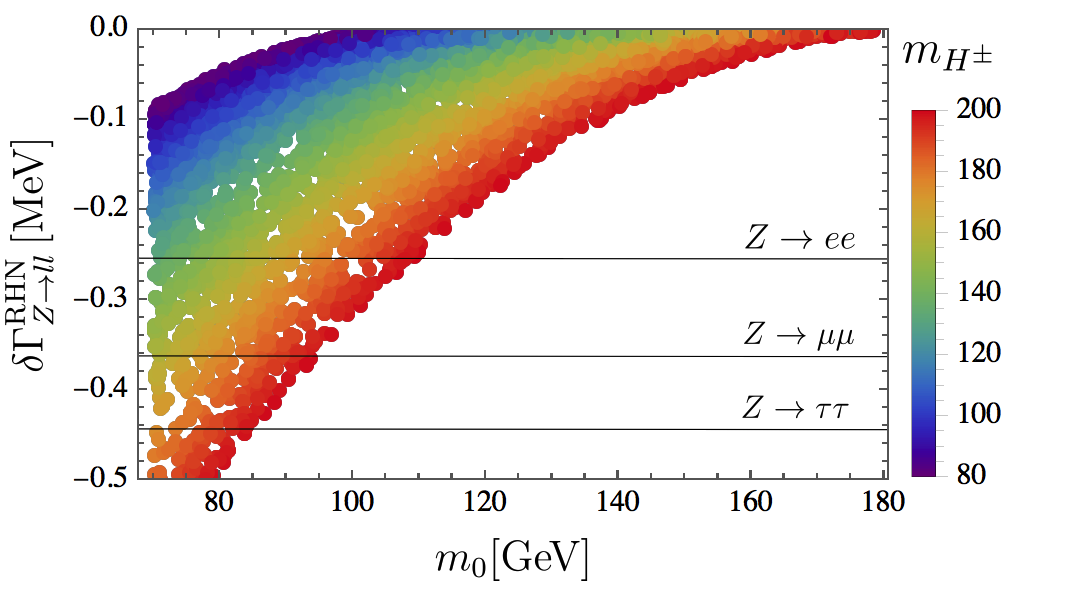}
\caption{\small Deviation of the decay rate of $Z\rightarrow l^+l^-$ as a function of the average scalar mass $m_0$ in Eq.~(\ref{eq:scalarmasses}) and the charged scalar mass $m_{H^\pm}$. The horizontal lines correspond to the bounds in Table~\ref{tab:contsraints} taken at $2\sigma$.}
\label{fig:dGammaZee}
\end{figure}
Fig.~\ref{fig:Zll_Znunu}. These are vertex corrections involving a combination of two inert scalars and a RHN inside the loop, which are absent in the exclusive IDM. The calculation of the necessary decay rates is detailed in Appendix \ref{app:Zll_Znunu} where we can safely neglect the lepton masses. In Fig.~\ref{fig:dGammaZee} we show the deviation of the decay rate $\delta\Gamma^{\rm RHN}_{Z\rightarrow ll}$ as a function of the average scalar mass $m_0$ in Eq.~(\ref{eq:scalarmasses}) and the charged scalar mass in the range $m_{H^\pm}\simeq \{70-180\}$~GeV. The horizontal lines correspond to the experimental bounds in Table~\ref{tab:contsraints} at $2\sigma$ where the most stringent one comes from $Z\rightarrow e^+e^-$. We observe that, at fixed $m_{H^\pm}$, the mass splitting between the neutral and charged scalar particles is constrained. In particular, the larger is $m_{H^{\pm}}$ the stronger is the bound. For example, at the largest mass considered $m_{H^\pm}\simeq 200$~GeV, the mass of the neutral scalar has to be $m_0\gtrsim 110$~GeV.

\section{Higgs-strahlung}
\label{sec:Higgsstrahlung}
The targeted S-matrix element is a function of kinematical variables, such as energy and momenta, as well as of the $Z$-polarization $\lambda$ and spin $s_{\pm}$ of the colliding leptons.  
From Lorentz invariance, the amplitude for the process $e^+(s_+) e^-(s_-) \rightarrow Z(\lambda) h$ can be decomposed in a suitable basis of indipendent matrix elements $\mathcal{M}_j$ 
\bea
\mathcal M\left(\lambda; s_+, s_-\right) &= \sum_j F_j \mathcal M_j\left(\lambda; s_+, s_-\right) \,,\\
\eea
with model-dependent scalar form factors $F_j$ \cite{Fleischer:1982af,Denner:1991kt}. The structure of the amplitudes $\mathcal M_j\left(\lambda; s_+, s_-\right)$ depends only on the process and the order of the perturbative expansion. In the unpolarized scenario we investigate, the explicit form of the spinors and polarization vectors can be entirely ignored by exploiting known computational maneuvering to sum over them.  
Turning to kinematics, a four-point scattering amplitude is described by two independent variables. In a frame-independent approach, these are usually chosen among the Mandelstam variables $S,T$ and $U$ connected by the relation $S+T+U=2m^2_e + m^2_Z+m^2_h$. In the center-of-mass system of the colliding lepton pair, it is instead convenient to pick the beam energy $E = \sqrt{S}/2$ ($S\geq m^2_Z+m^2_h$) and the cosine of the angle $\theta$ between the $e^+$ and the $Z$. The relation between the Mandelstam variables and $\cos\theta$ is 
\bea
2S\beta_{S}\cos\theta = T-U,
\eea
where $\beta_{S}$ is the usual two-particle phase space function that, neglecting the electron mass, reads 
\bea
\beta_{S} = \frac{\sqrt{(m_h^2-m_Z^2)^2 - 2 (m_h^2 + m_Z^2) S + S^2}}{S}.
\eea
In terms of these, and confining ourselves to the unpolarized case, 
the differential cross-section is connected to the amplitude $\mathcal M\left(\lambda; s_+, s_-\right)$ via 
\bea
\label{eq:sigmaM}
\frac{d \sigma}{d \cos\theta} = \frac{\beta_{S}}{32 \pi S} \left(\frac{1}{4} \sum_{\lambda,s_+, s_-} |\mathcal M\left(\lambda; s_+, s_-\right)|^2\right).
\eea
For the calculation of $|\mathcal{M}|^2$ in Eq.~\ref{eq:sigmaM} at the Next to Leading Order (NLO) we have to compute \cite{Denner:1991kt}
\bea
|\mathcal{M}^{\rm NLO}|^2 \simeq
\sum_{ij} (F^{\rm LO}_i)^* (F^{\rm LO}_j +2 \delta F_j){\rm Re}[\mathcal{M}^*_i \mathcal{M}_j],
\eea
where we denote $F^{\rm LO}$ and $\delta F_j$ the LO and the NLO form factors, respectively. If the NLO correction takes place at one-loop order, the latter are linear combinations of the scalar Passarino-Veltman (PV) integrals \cite{Passarino:1978jh,tHooft:1978jhc}. In the following, we adopt the conventions for the PV-integrals of \cite{Hahn:2000jm,Hahn:2000kx,Hahn:2016ebn}.  

Analogous to the LEP experience, to reveal the presence of new particles not involved in the tree-level exchange, a coherent inclusion of their radiative effects is required. Such technical operation demands a regularization prescription and a renormalization scheme. While the first is almost always chosen to be dimensional regularization, for the second, and the nature of the process under study, the literature offers different competing options \cite{Sirlin:1980nh,Passarino:1989ta,Novikov:1993ic,Marciano:1980be,Degrassi:1990tu}. We opted for an On-Shell (OS) prescription, and sketch in Appendix \ref{OSRenoApp} the main steps of the procedure while referring to \cite{Denner:1991kt,Denner:1992bc,Abouabid:2020eik} for a more detailed analysis.  
\begin{figure}[t]
\centering
\includegraphics[scale=0.33]{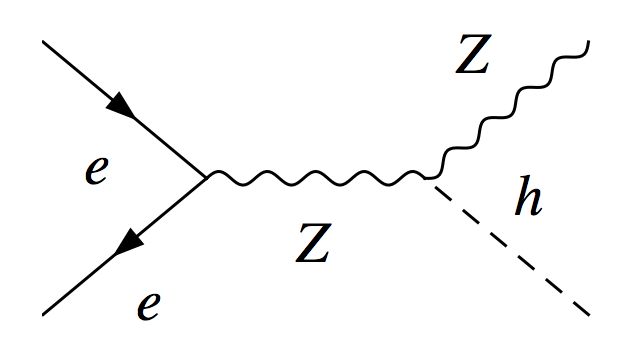}
\caption{The dominant contribution to $e^+ e^- \rightarrow Z h$ in the SM. }
\label{fig:eeZHLO}
\end{figure}
In the SM, the scattering process $e^+e^-\rightarrow Zh$ 
happens at LO through the tree-level exchange of a $Z$ boson in Fig.~\ref{fig:eeZHLO}.  This represents the dominant contribution since we can safely neglect the vertex $e^+e^- h(G^0)$ proportional to the tiny electron mass. The LO amplitude can be expanded in terms of the two forms 
\bea 
\label{eq:HelBasis}
& \mathcal M_L\left(\lambda; s_+, s_-\right) = \bar v(s_+)  \slashed{\epsilon}^*(\lambda) P_L u(s_-)\, , \\
& \mathcal M_R\left(\lambda; s_+, s_-\right) = \bar v(s_+) \slashed{\epsilon}^*(\lambda) P_R u(s_-) \, , 
\eea
where the chirality projectors are $P_{L,R} = \left( 1 \mp \gamma_5\right)/2$. The LO form factors read  
\bea
\label{eq:CLRLO}
F^{\rm LO}_{L,R} = \frac{4 \alpha \pi\, m_Z\,g^{Ze}_{L,R}}{c_Ws_W(S-m^2_Z)},
\eea
where 
\bea
\label{eq:geLR}
g^{Ze}_L=\frac{s^2_W-\frac{1}{2}}{c_Ws_W}  \quad \text{and} \quad g^{Ze}_R=\frac{s_W}{c_W}
\eea
are the $Z$-boson couplings to $e_L$ and $e_R$, $\alpha$ is the usual fine structure constant and $s_W$, $c_W$ are the sine, cosine of the Weinberg angle. In the OS scheme, these are, at all orders in the perturbation expansion, shortcuts for the expressions $c_W = m_W/m_Z$ and $s_W^2 =1 - m_W^2/m^2_Z$.
Plugging Eq.~(\ref{eq:CLRLO}) into Eq.~(\ref{eq:sigmaM}), and using that
\bea
\frac{1}{4}\sum_{\lambda;s_+,s_-}|\mathcal{M}_{L,R}|^2=\frac{S}{2}\left[1+\frac{S\beta^2_S}{8 m^2_Z}(1-\cos\theta^2)\right],
\eea
upon integration over $\theta$, we arrive at the formula
\bea 
\label{eq:sigmaZhLO}
 &\sigma^{\rm LO}_{Zh} (S) = \frac{\beta_{S}}{32\pi} \left(1 + \frac{S}{12} \,\frac{\beta^2_{S}}{m_Z^2}\right) \bigg[ |F^{\rm LO}_L|^2+|F^{\rm LO}_R|^2\bigg] .
\eea

\begin{figure}[t]
\centering
\includegraphics[scale=0.6]{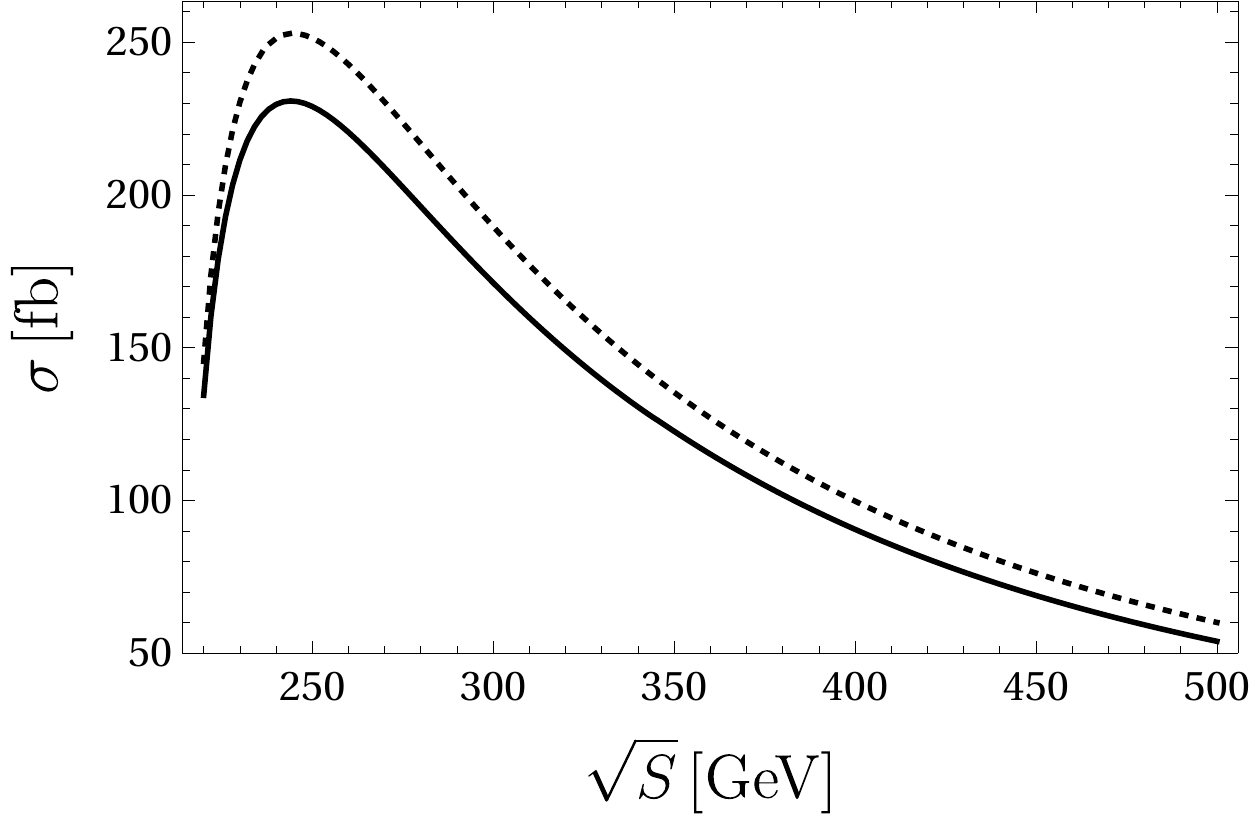}
\includegraphics[scale=0.6]{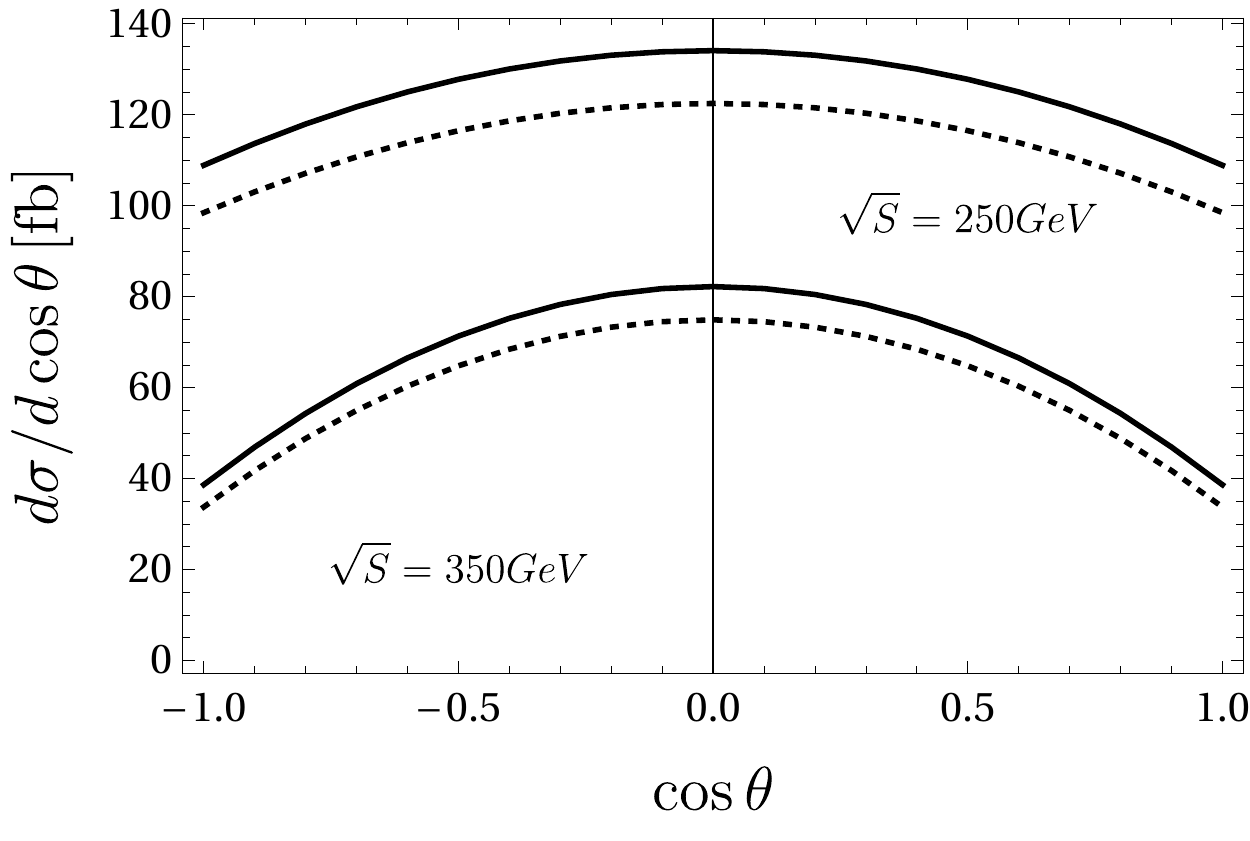}
\caption{\small Top: comparison of energy behaviours for the unpolarized SM cross-sections for $e^+ e^- \rightarrow Z h$. Leading order (dotted line) vs SM at one-loop (solid line). Bottom: comparison of angular distributions at two different values of $\sqrt{S}$.}
\label{fig:SMXS}
\end{figure}
Here $F^{\rm LO}_L$ gives a slightly bigger contribution than $F^{\rm LO}_R$, with $(F^{\rm LO}_L/F^{\rm LO}_R)^2 \simeq 1.54$.
The cross-section of the leading unpolarized SM peaks at around $E \sim 244$ GeV for a value of $\sim 254$ fb. Moreover, the longitudinal component of the $Z$ dominates the cross-section, softening its high-energy behaviour \cite{Denner:1991kt,Fleischer:1982af}.
In light of the projected energy resolution, an assessment of the QED/EW radiative corrections 
has been the object of several dedicated studies. Before focusing on the narrower scenario needed for our work, the profiling of radiative corrections proper of the Scotogenic model, we briefly summarize the main results and peculiarities of the SM computation. 
The major distinction arises from dealing with IR divergences and their cancellation from the final cross-section. Triangle diagrams with propagating photons provide loop divergent results regularized by a fictitious photon mass $m_{\gamma}$. As known, the disappearance of the infrared regulator from the final cross-section is ensured by a proper inclusion of soft photon radiation, of energy $E_{soft} < \Delta E \ll \sqrt{S}$. The contribution of soft photons simply factorizes as \cite{Denner:1991kt,Denner:1992bc}
\bea \label{brstr0}
&\frac{d \sigma^{soft}_{Zh\gamma}}{d \cos\theta} = \frac{d \sigma^{\rm LO}_{Zh}}{d \cos \theta} \delta^{soft}\, ,
\eea
where
\bea \label{brstr1}
&\delta^{soft} = \\
&-\frac{\alpha}{\pi} \left[ \ln \frac{(2 \Delta E)^2}{m_{\gamma}^2}\bigg(1 + \ln \frac{m_e^2}{S}\right) + \frac{1}{2}\ln^2 \frac{m_e^2}{S} + \ln \frac{m_e^2}{S} + \frac{\pi^2}{3}  \bigg] .  \\
\eea
After the inclusion of \ref{brstr0}, the remaining QED/EW one-loop corrections only introduce divergences of UV nature. These can be studied within the same renormalization framework we adopted to deal with states proper of the Scotogenic model. Before presenting the details of the UV-renormalization, we gather in Fig.~\ref{fig:SMXS} the full one-loop Standard-Model prediction for $e^+ e^- \rightarrow Z h$. In the top panel, we show the unpolarized LO cross-section in Eq.~\ref{eq:sigmaZhLO} as a function of $\sqrt{S}$ (dotted line) and compare it with the full NLO calculation in the SM (solid line).
We notice the importance of the inclusion of the QED/EW NLO corrections, accounting for a reduction of $\sim 25$ fb around the production peak. In particular, as SM benchmark points used to assess the NP contribution, we use the two reference values  
\bea
\label{eq:sigmaSM_250_350}
&\sigma^{\rm SM}_{Zh}(250 \text{GeV})=228.748\text{[fb]},\\  
&\sigma^{\rm SM}_{Zh}(350 \text{GeV})=123.392\text{[fb]},
\eea
from Ref.~\cite{Abouabid:2020eik}.
In the bottom panel, we fix the energy at these two representative values of $\sqrt{S}$ and plot the differential cross-section as a function of $\cos\theta$.
Higher order corrections, as well as the onset of NP, will, in general, introduce new terms beyond the tree-level ones of \ref{eq:HelBasis}, with effects that can be revealed by dedicated scrutiny of the polarized cross-section.  
We leave this, and in particular, the interesting effects of CP violation over the helicity structure \cite{Gounaris:1995mx,Gounaris:2014tha,Grzadkowski:1999ye}, to a future study.
As discussed in sec.\ref{sec:Model}, the Scotogenic model introduces extra sterile RHN and a $3 \times 3$ complex Yukawa matrix on top of the IDM setup. Thus, the full Higgstrahlung computation in the Scotogenic model encompasses three stages, having to deal first with the EW-contributions of the SM, then with the extra doublet of the IDM, and finally with the virtual propagation of the RHNs. For some of the counterterms in \ref{CoRen} and \ref{FiRen}, NP contributions are limited to those of the IDM and their form can therefore be found in past studies such as \cite{Abouabid:2020eik}. A subset of the counterterms is instead modified by the virtual propagation of RHNs states and we provide the corresponding one-loop formulas in the appendix \ref{appCT}. 
Finally, we stress that, at the perturbative order needed, and considering that only SM particles inhabit the external states, we can ignore the renormalization of couplings proper of the Scotogenic sector.

For the model implementation, we used the FeynRules and FeynArts packages. The computation of the sets of one-loop corrections has been performed with the help of the FormCalc package and LoopTools for the evaluation of the PV-functions. 

\subsection{IDM contributions}
\label{sec:IDMcontributions}
\begin{figure}[t]
\centering
\includegraphics[scale=0.16]{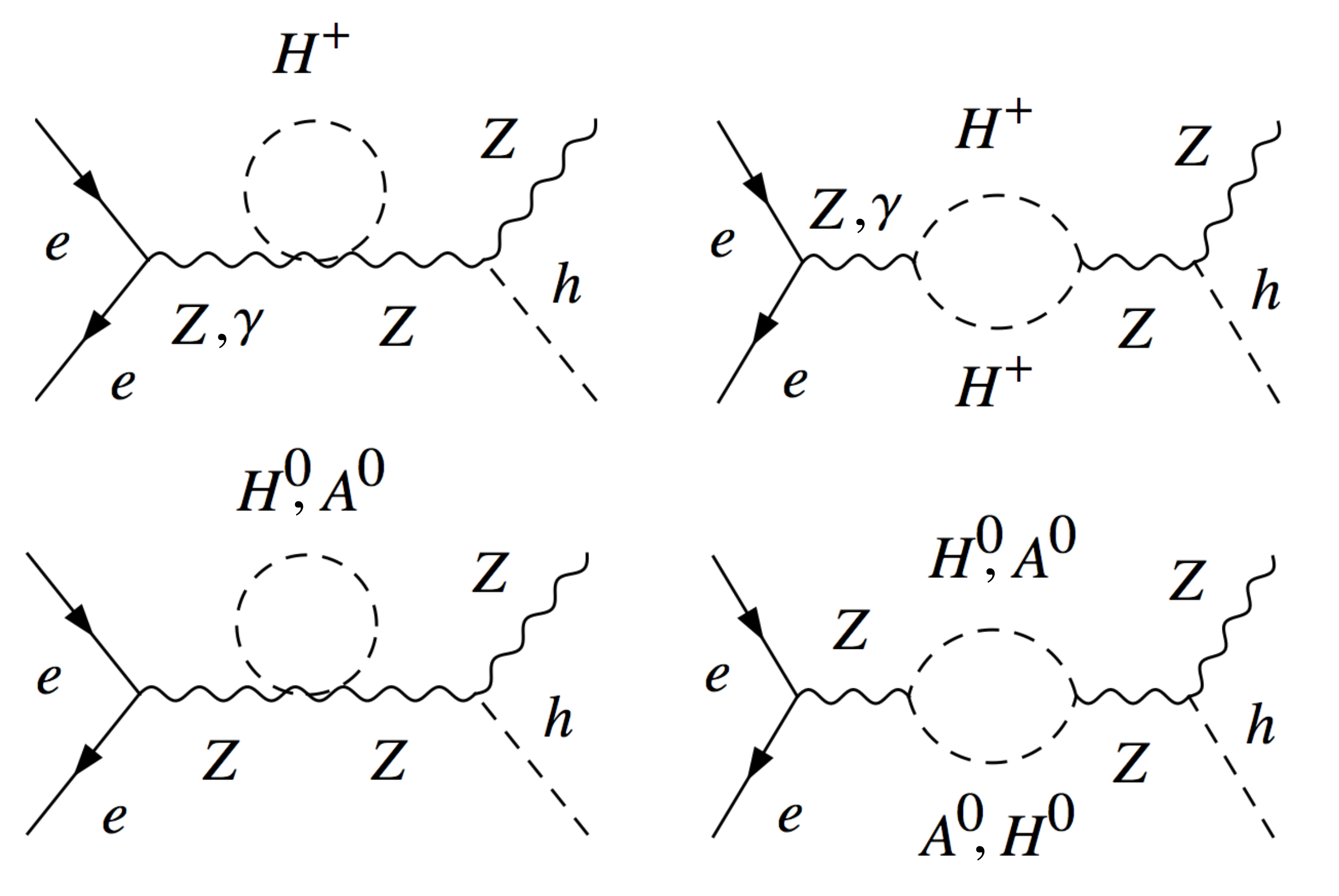}
\caption{\small Dominant self-energy contributions to $e^-e^+ \rightarrow Zh$ in the IDM model.}
\label{fig:selfenergiesIDM}
\includegraphics[scale=0.16]{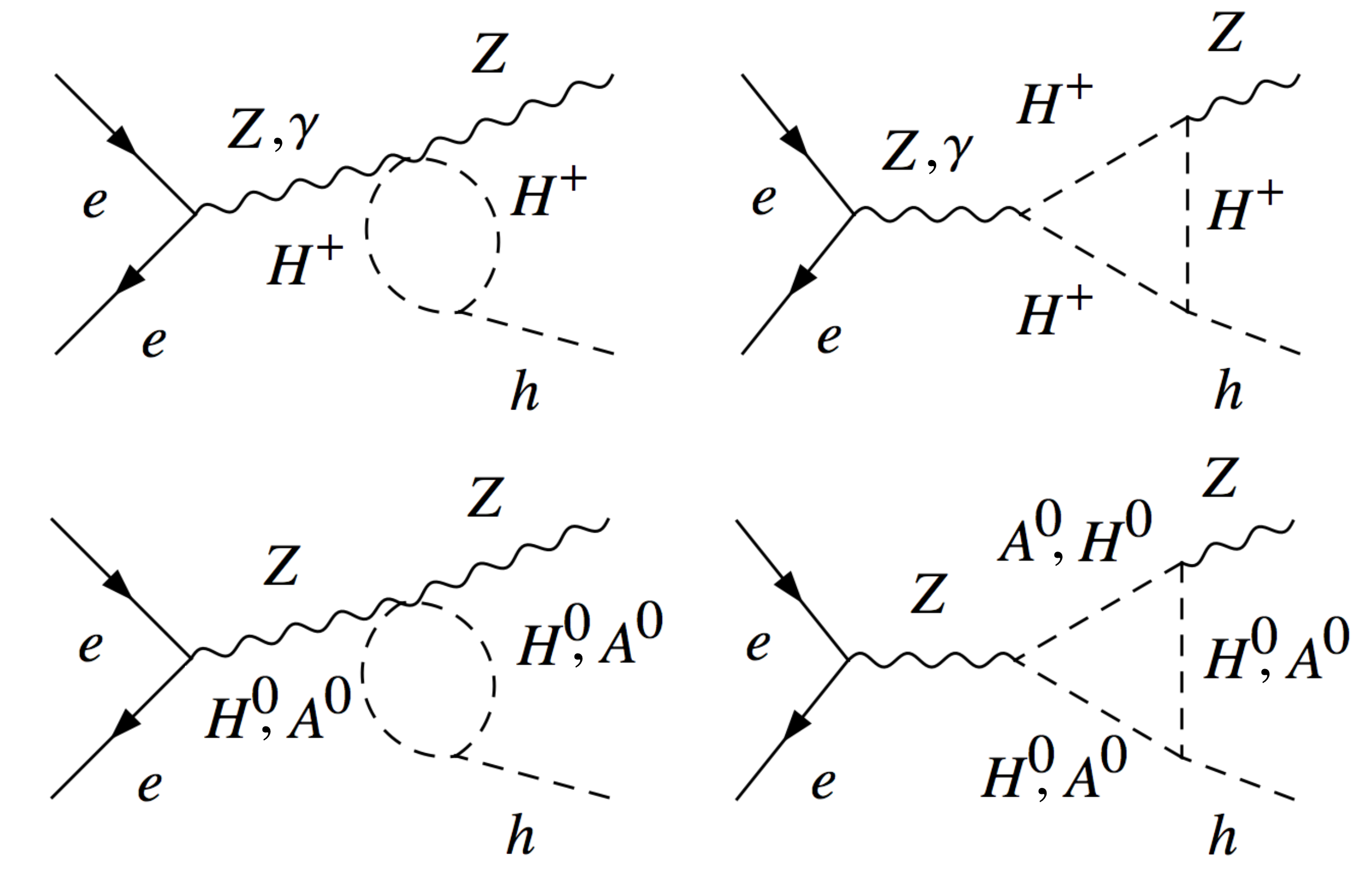}
\caption{\small Dominant vertex diagram contributions to $e^-e^+ \rightarrow Zh$ in the IDM model.}
\label{fig:trianglesIDM}
\end{figure}

\begin{figure}[t]
\centering
\includegraphics[scale=0.18]{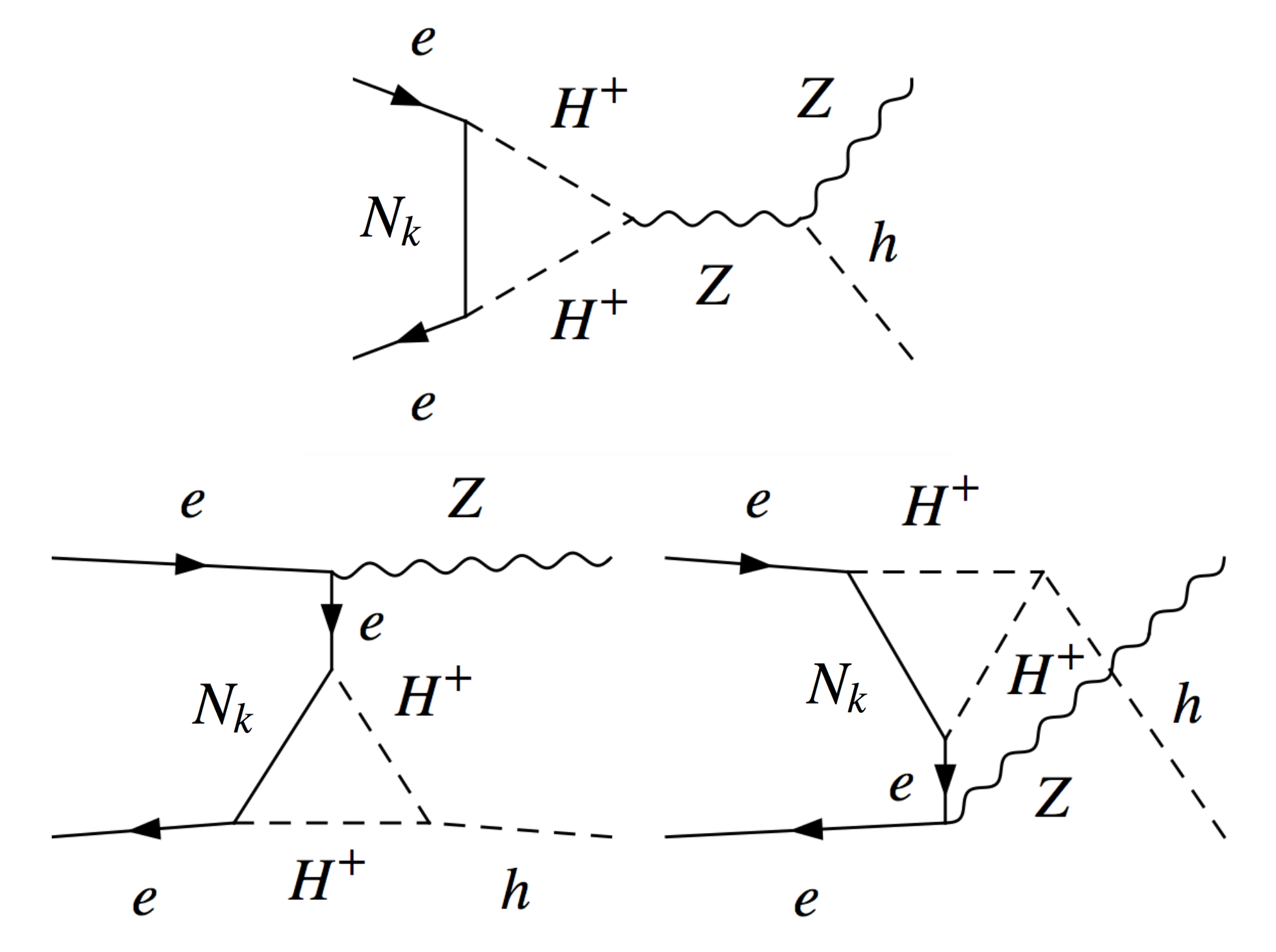}
\caption{\small Dominant vertex diagram contributions to $e^-e^+ \rightarrow Zh$ introduced by the RHN interactions of the Scotogenic model.}
\label{fig:triangles}
\includegraphics[scale=0.18]{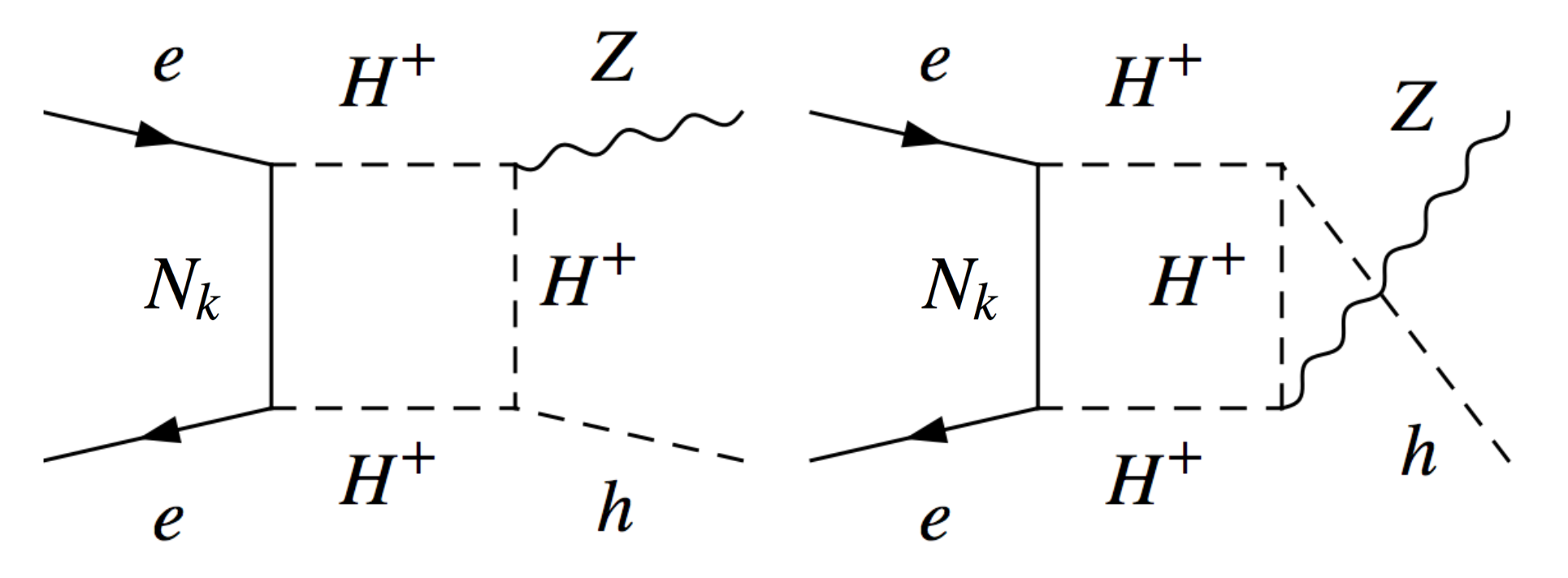}
\caption{\small Dominant box diagram contributions to $e^-e^+ \rightarrow Zh$ introduced by the RHN interactions of the Scotogenic model.}
\label{fig:boxes}
\end{figure}
The radiative corrections proper of the IDM are depicted in Fig.~\ref{fig:selfenergiesIDM} and Fig.~\ref{fig:trianglesIDM}. They split into self-energy corrections to the $Z$ propagator and vertex corrections to the $hZZ$-coupling. Notice that, as far as the additional doublet is inert, i.e. it does not couple to the SM fermionic fields, no box topologies involving the new scalars can be generated. The self-energies of Fig.~\ref{fig:selfenergiesIDM} correct the coefficient of both $\mathcal M_{L,R} \left(\lambda; s_+, s_-\right)$ in Eq.~(\ref{eq:HelBasis}):
\bea
\label{eq:IDMself}
\delta F^{S,\rm IDM}_{L,R} &=\frac{\alpha F^{\rm LO}_{L,R}}{8\pi c^2_W s^2_W (S-m_Z^2)}\\
\times \bigg\{& (c^2_W-s^2_W)^2 \bigg[1-\frac{2c_Ws_W(1-m^2_Z/S)}{g^{Ze}_{L,R}(c^2_W-s^2_W)}\bigg]\\
&\times \bigg(2 B_{00} [S,m_{H^\pm},m_{H^\pm}]-A_0[m_{H^\pm}]\bigg)\\
&+\bigg(2 B_{00} [S,m_{A^0},m_{H^0}]-\frac{A_0[m_{H^0}]+A_0[m_{A^0}]}{2}\bigg)\bigg\} \,,
\eea
\bea
(\delta F^{S,\rm IDM}_{L,R})^{ct} = &F^{\rm LO}_{L,R}\bigg[
\frac{\delta m^2_{Z}}{S-m_Z^2}-\delta_{Z}- \frac{\delta_{AZ}}{2g^{Ze}_{L,R}} \\
 &+\frac{\delta_{ZA}}{2g^{Ze}_{L,R}} \bigg(1-\frac{m_Z^2}{S}\bigg)\bigg].
\eea
where clearly the $A_0$ functions correspond to the tadpole topologies (left diagrams) and the $B_{00}$ to the bubble topologies (right diagrams).
The vertex corrections in Fig.~\ref{fig:trianglesIDM} give another contribution to $\mathcal{M}_{R,L}$ given by
\bea
\label{eq:IDMvert}
\delta F^{V, \rm IDM }_{L,R} & = \frac{F^{\rm LO}_{L,R}}{16\pi^2} \bigg\{(c^2_W-s^2_W)^2\bigg[ 1-\frac{2c_Ws_W(1-m^2_Z/S)}{ g^{Ze}_{L,R}(c^2_W-s^2_W)}\bigg]\\
&\times \lambda_3 \bigg(-B_0[m_h^2,m^2_{H^\pm},m^2_{H^\pm}] \\
&\hspace{0.7cm}+4C_{00}[m^2_Z,m_h^2,S,m^2_{H^\pm},m^2_{H^\pm},m^2_{H^\pm}] \bigg) \\
&+\frac{\lambda_3+\lambda_4 + \lambda_5}{2} \bigg(-B_0[m_h^2,m^2_{H^0},m^2_{H^0}] \\
&\hspace{0.7cm}+4C_{00}[m^2_Z,m_h^2,S,m^2_{A^0},m^2_{H^0},m^2_{H^0}]\bigg)\\
&+\frac{\lambda_3+\lambda_4 - \lambda_5}{2} \bigg(-B_0[m_h^2,m^2_{A^0},m^2_{A^0}] \\
&\hspace{0.7cm} +4C_{00}[m^2_Z,m_h^2,S,m^2_{H^0},m^2_{A^0},m^2_{A^0}] \bigg)\bigg\},
\eea
\bea
(\delta F^{V, \rm IDM }_{L,R})^{ct}& =F^{\rm LO}_{L,R}\bigg[ \frac{\delta_h}{2} + \delta_Z + \delta{\rm e} + \frac{\delta m^2_Z}{2m^2_Z} \\
& + \frac{c^2_W-s^2_W}{2s^2_W}\bigg( \frac{\delta m^2_Z}{m^2_Z}-\frac{\delta m^2_W}{m^2_W}\bigg)\bigg]
\eea
where the $B_0$ and $C_{00}$ functions arise from the bubble topologies (left diagrams) and triangle topologies (right diagrams) correspondingly.
The diagrams in Fig.~\ref{fig:trianglesIDM} involve two gauge couplings and a quartic coupling, which is $\lambda_3$ for the two top diagrams and the combinations $\lambda_3 +\lambda_4 \pm \lambda_5$ for the bottom diagrams.  An interesting remark concerns the fact that the UV divergent parts of the right and left diagrams in \ref{fig:trianglesIDM} cancel between each other such that the IDM vertex diagrams are eventually UV finite. However, the consistency of the On-shell scheme requires a finite counterterm given in the appendix.
In the Scotogenic scenario where $\lambda_5\simeq 0$ these formulas further simplify setting $m_{H^0}\simeq m_{A^0} \simeq m_0$ so that the last two lines in (\ref{eq:IDMvert}) are equal. 
Moreover, a finite value is generated for the coefficients of 
\bea 
\label{eq:HelBasisBox}
& \mathcal{M}_{L,R,h}\left(\lambda; s_+, s_-\right) = \bar v(s_+)  \slashed{k}_h P_{L,R}  u(s_-)\epsilon^*(\lambda)\cdot k_{h} \, , 
\eea
equal to 
\bea
\delta F^{V, \rm IDM }_{L,R,h}  &= -\frac{F^{\rm LO}_{L,R}}{4\pi^2}\bigg\{(c^2_W-s^2_W)^2\bigg[ 1-\frac{2c_Ws_W(1-m^2_Z/S)}{g^{Ze}_{L,R}(c^2_W-s^2_W)}\bigg]\\
& \times \lambda_3 \bigg(C_{2}[m^2_Z,m_h^2,S,m^2_{H^\pm},m^2_{H^\pm},m^2_{H^\pm}]\\
&\hspace{0.7cm}+C_{12}[m^2_Z,m_h^2,S,m^2_{H^\pm},m^2_{H^\pm},m^2_{H^\pm}]\\
&\hspace{0.7cm}+C_{22}[m^2_Z,m_h^2,S,m^2_{H^\pm},m^2_{H^\pm},m^2_{H^\pm}] \bigg) \\
&-\frac{\lambda_3+\lambda_4+\lambda_5}{2} 
\bigg(C_{12}[m^2_Z,m_h^2,S,m^2_{A^0},m^2_{H^0},m^2_{H^0}] \bigg)\\
&+\frac{\lambda_3+\lambda_4-\lambda_5}{2}\bigg(C_{1}[m^2_Z,m_h^2,S,m^2_{H^0},m^2_{A^0},m^2_{A^0}]\\
&\hspace{0.7cm}+C_{11}[m^2_Z,m_h^2,S,m^2_{H^0},m^2_{A^0},m^2_{A^0}]\\
&\hspace{0.7cm}+C_{12}[m^2_Z,m_h^2,S,m^2_{H^0},m^2_{A^0},m^2_{A^0}] \bigg)
\bigg\}.
\eea
\subsection{RHN contributions}
\label{sec:RHNcontributions}
Here we focus on the contributions that are unique to the Scotogenic case, that is on those concerning the Yukawa-dependent non-universal corrections. 
Being gauge singlets, the RHNs do not couple to the $Z$-boson so in this case, exclusively vertex and box topologies of the corrections are present. These are shown in figs.~\ref{fig:triangles} and \ref{fig:boxes}, respectively, where a copy of them is implied for each one of the three RHNs ($k=1,2,3$). These diagrams involve only a subset of the entries of the full Yukawa matrix $Y_N$, that is the first row of the Yukawa couplings of the RHN to the electron: $y^N_{1k}$. Notice that the neutral scalars $H^0$ and $A^0$ do not intervene in these contributions and they depend uniquely on the mass of the new charged scalar $H^\pm$.
The computation of the vertex corrections depicted in Fig.~\ref{fig:triangles} produces extra polarization amplitudes compared to the ones in \ref{eq:HelBasis}. While we have maintained a non-zero electron and muon masses over the full renormalization, when we consider the massless limit, the triangles \ref{fig:triangles} simplify and only correct the coefficient of $\mathcal M_L\left(\lambda; s_+, s_-\right)$. This simpler formula is 
\bea \label{ScotoTri}
& \delta F_L^{V,\rm RHN} = -\frac{ m_{Z}(c^2_W-s_W^2)}{16 \pi^2} \sum^3_{k=1}  |y^N_{1k}|^2 \\
&\times \bigg( \frac{ 4\pi\alpha}{c_W^2 s_W^2 (S-m_Z^2)} C_{00}\left[S, 0, 0, m^2_{H^\pm},m^2_{H^\pm}, m_{N_k}^2\right]  \\
& \hspace{1cm}- \lambda_3\,C_{1}\left[m^2_h, T, 0, m^2_{H^\pm},m^2_{H^\pm}, m_{N_k}^2\right]  \\
& \hspace{1cm}+ \lambda_3\,C_{1}\left[m^2_h, U, 0, m^2_{H^\pm},m^2_{H^\pm}, m_{N_k}^2\right]  \bigg) , \\
\eea
 \bea
 (\delta F_{L,R}^{V,\rm RHN})^{ct} = &F^{\rm LO}_L \bigg[ \frac{\delta_Z}{2}+\delta_{e_L} +\delta{\rm e}+ \frac{\delta_{AZ}}{2 g^{Ze}_L}\\
 &+\frac{1}{2 s^2_W}\bigg( \frac{\delta m^2_W}{m^2_W} - \frac{\delta m^2_Z}{m^2_Z}\bigg) \bigg]
 \eea
where the three $C$ scalar functions are in one-to-one correspondence with the diagrams in Fig.~\ref{fig:triangles}.
In the same approximation of zero electron mass, also the boxes Fig.~\ref{fig:boxes} arrange in a simpler contribution over $\mathcal M_L\left(\lambda; s_+, s_-\right)$ :
\bea
& \delta F_L^{B,\rm RHN} = - \lambda_3 \frac{m_Z(c^2_W-s_W^2) }{8 \pi^2} \sum^3_{k=1}  |y^N_{1k}|^2 \\
&\times  \bigg( D_{00}\left[m^2_h,m^2_Z, 0, 0, S, U, m^2_{H^\pm},m^2_{H^\pm},m^2_{H^\pm},  m_{N_k}^2\right]  \\
& \hspace{0.2cm}+ D_{00}\left[m^2_Z,m^2_h, 0, 0, S, T, m^2_{H^\pm},m^2_{H^\pm},m^2_{H^\pm},  m_{N_k}^2\right] \bigg) \, . \\
\eea
Moreover, a finite value is generated for the coefficients of the two helicity amplitudes   
\bea \label{HelBasisBox}
& \mathcal{M}_{L,e}\left(\lambda; s_+, s_-\right) = \bar v(s_+)  \slashed{k}_h P_L  u(s_-) \epsilon^*(\lambda)\cdot k_{e^-} \, , \\ 
& \mathcal{M}_{L,h}\left(\lambda; s_+, s_-\right) = \bar v(s_+)  \slashed{k}_h P_L  u(s_-) \epsilon^*(\lambda)\cdot k_{h} \, , 
\eea
equal to 
\bea
& \delta F_{L,e}^{B,\rm RHN} = - \lambda_3 \frac{m_Z(c^2_W-s_W^2) }{8 \pi^2}  \sum^3_{k=1}  |y^N_{1k}|^2 \\
&\times \bigg(D_{13}\left[m^2_h,m^2_Z, 0, 0, S, U, m^2_{H^\pm},m^2_{H^\pm},m^2_{H^\pm}, m_{N_k}^2\right]  \\ 
& \hspace{0.2cm}- D_{13}\left[m^2_Z,m^2_h, 0, 0, S, T, m^2_{H^\pm},m^2_{H^\pm},m^2_{H^\pm},  m_{N_k}^2\right] \bigg) \, , 
\eea
and 
\bea
& \delta F_{L,h}^{B,\rm RHN} = - \lambda_3 \frac{m_Z(c^2_W-s_W^2) }{8 \pi^2} \sum^3_{k=1}  |y^N_{1k}|^2 \\
& \times \bigg(D_{1}\left[m^2_h,m^2_Z, 0, 0, S, U, m^2_{H^\pm},m^2_{H^\pm},m^2_{H^\pm},  m_{N_k}^2\right]  \\ 
& \hspace{0.2cm}+ D_{11}\left[m^2_h,m^2_Z, 0, 0, S, U, m^2_{H^\pm},m^2_{H^\pm},m^2_{H^\pm},  m_{N_k}^2\right]  \\ 
& \hspace{0.2cm}+ D_{12}\left[m^2_h,m^2_Z, 0, 0, S, U, m^2_{H^\pm},m^2_{H^\pm},m^2_{H^\pm},  m_{N_k}^2\right]   \\ 
& \hspace{0.2cm}- D_{12}\left[m^2_Z,m^2_h, 0, 0, S, T, m^2_{H^\pm},m^2_{H^\pm},m^2_{H^\pm},  m_{N_k}^2\right] \bigg)\, . 
\eea

The renormalization of the UV-divergences generated by $\delta F_L^{V,\rm RHN}$ can be confronted with the OS recipe illustrated in Appendix \ref{OSRenoApp}. Again, for completeness and self-consistency, we have computed the renormalization of the full Scotogenic model, including the SM sector. Nevertheless, confining ourselves to the non-universal corrections, only the computation of $\delta m_f$ and $\delta \psi_{L,R}$ is needed \ref{appCT}.

\section{Results}
\label{sec:results}
The inclusion of the theoretical and phenomenological bounds has a strong impact over the same parameter space that can favourably produce a collider signal at FCC-ee. 
It is the main message of this work that regions with interesting signatures survive, therefore illuminating an interesting link between neutrino physics and forthcoming collider phenomenology.  
\begin{table}[h]
\centering
{\renewcommand{\arraystretch}{1.5} 
\resizebox{1\columnwidth}{!}{
\begin{tabular}{c c c}
\hline
$\alpha(m_e) =1./137.035999$ &  $\alpha(m_Z) = 1./128.943$  & \\
$m_Z = 91.1876$ GeV & $m_W = 80.379$ GeV & $ m_h = 125.18$ GeV \\ 
$m_e = 0.000511$ GeV & $m_{\mu} = 0.10$ GeV & $ m_{\tau} = 1.74$ GeV \\
$m_u = 2.16 \cdot 10^{-3}$ GeV  & $m_c = 1.27$ GeV   & $ m_t = 172.96  $ GeV \\ 
$m_d = 4.67 \cdot 10^{-3}$ GeV & $m_s = 93 \cdot 10^{-2}$ GeV & $ m_b = 4.18$ GeV  \\
\hline
\end{tabular}}
\caption{\small Values of the SM parameters used in this paper, taken from \cite{Workman:2022ynf}}
\label{tab:SMInputs}
}
\end{table}
The first step in our analysis consists in finding the conditions to enhance those features proper of the Scotogenic model. In generic extensions with RHN states, this translates into adopting sizable values of the extra Yukawa terms and, indeed, we find that interesting collider signatures can be accessed by this straightforward selection of the parameter space. The connection between the new Yukawa sector and neutrino phenomenology, behind the introduction of the Scotogenic model, violently constrains the freedom in admitting large Yukawas and simultaneously avoiding a too-large scale for NP. The clearest consequence is that we must enforce a strong degeneracy between the neutral components of the inert doublet \ref{eq:mnuExp}, generating an almost vanishing $\lambda_5$. It is a happy circumstance that this can be naturally achieved in the Scotogenic model, where small values of $\lambda_5$ can be linked to the breaking of a $U(1)$ symmetry of Peccei-Quinn type \cite{Dasgupta:2013cwa,Suematsu:2017kcu}, thus being protected from large radiative corrections. On top of the extra Yukawa, a survey of the diagrams \ref{fig:triangles}-\ref{fig:boxes} points to the inert charged current as necessarily non-zero, in order to have a visible Scotogenic contribution. Among the diagrams in \ref{fig:triangles} we find that only one can be non-zero in a scenario with $\lambda_3 = 0$, a scenario connected to all the IDM scalars being degenerate in mass. We have explicitly explored this case and found it unable of providing visible signatures, beyond those proper of the IDM, at the required experimental resolution. Therefore we favour large, but still perturbative, $\lambda_3$ and consider masses of the charged scalar $m_{H^{\pm}}$ not too far from their lower bounds. 

The scan we conducted in order to highlight the discovery potential of the new states is split into different stages. To start, a scan within the IDM parameters is performed in order to select points in the space of $\lambda_2, \lambda_3, m_{H_0}, m_{A_0}$ and $, m_{H^{\pm}}$, which survive the main theoretical and phenomenological bounds of section~\ref{sec:Model}. Such a preliminary scan is realized within the values of \ref{tab:scan0} 
\begin{table}[h!]
\centering
{\renewcommand{\arraystretch}{1.5} 
\resizebox{0.9\columnwidth}{!}{
\begin{tabular}{c c}
\hline
$m_{H^{\pm},H^0} = \left\{80, 200\right\}$ GeV $\qquad \qquad$& $\lambda_2 = \left\{0, 4\pi/3\right\}$  \\
$ m_{H_0} - m_{A_0} = {\left\{10^{-9}, 10^{-7}\right\}}$ GeV $\quad$  & $\lambda_3 = \left\{-1.49, 1.4\right\}$ \\
\hline
\end{tabular}}
\caption{\small The range of the scan for points in the parameter space of the IDM.}
\label{tab:scan0}
}
\end{table}
and, for each of these points, the corresponding $e^- e^+ \rightarrow Zh$ cross-section is also computed. 

The surviving set will seed a second scan which will instead, for each IDM scenario represented by them, thoroughly explore the space of Yukawa couplings and RH neutrino masses. This is initiated with the selected values of $m_{H_0},  m_{A_0}$, and a randomly selected value of $m_{N_{i}}$ in the interval $\left\{0.01, 2\right\} m_{H^{\pm}}$. These are the necessary inputs we use to define $\Lambda_k$ in \ref{eq:mnu}.
To complete the random generation of $Y_N$ a more involved procedure is required, intertwined with the Casas-Ibarra procedure \ref{sec:NeuSection} so that different random variables are called for to comply with neutrino masses and mixing in either the NH or the IH setup. 
We found that the non-zero sizable Yukawa are subjected to the strongest constraints, in particular coming from the tight bounds of lepton flavour violation discussed in section \ref{sec:NeuSection}. To optimize our exploration we use the freedom given by the complex orthogonal matrix $R$ in \ref{eq:CasasIbarra}, to automatically impose, when possible, the  BR$(\mu \rightarrow e \gamma)$ bound and enforce towards values of forthcoming experimental probing.
The remaining inputs required by the Casas-Ibarra ansatz are given by the generation of the PMSN matrix, which we perform by picking within the values of \ref{tab:neutrinoData}, and a choice for the scale of the lightest neutrino. For the latter we accommodate for the broad range $\left\{10^{-7}, 10^{-1}\right\}$ eV.
No other impositions are made and the generated $m_{N_i}$ and $Y_N$ are confronted with all the bounds discussed previously as well as their leptonic anomalous magnetic moments. 
Finally, for such points of the full Scotogenic parameter space, we produce the $e^- e^+ \rightarrow Zh$ cross-section. The main target of our analysis will be the assessment of how much of the full $\sigma_{Zh}$ can be blamed on the presence of the extra RHN states as to positively resolve, via the precise FCC-ee measurement, the discovery of the Scotogenic setup over the dominant IDM background. 

\subsection{Numerical Analysis: Generic Results}
\label{sec:Analysis}
\begin{figure}[t]
\centering
\includegraphics[scale=0.11]{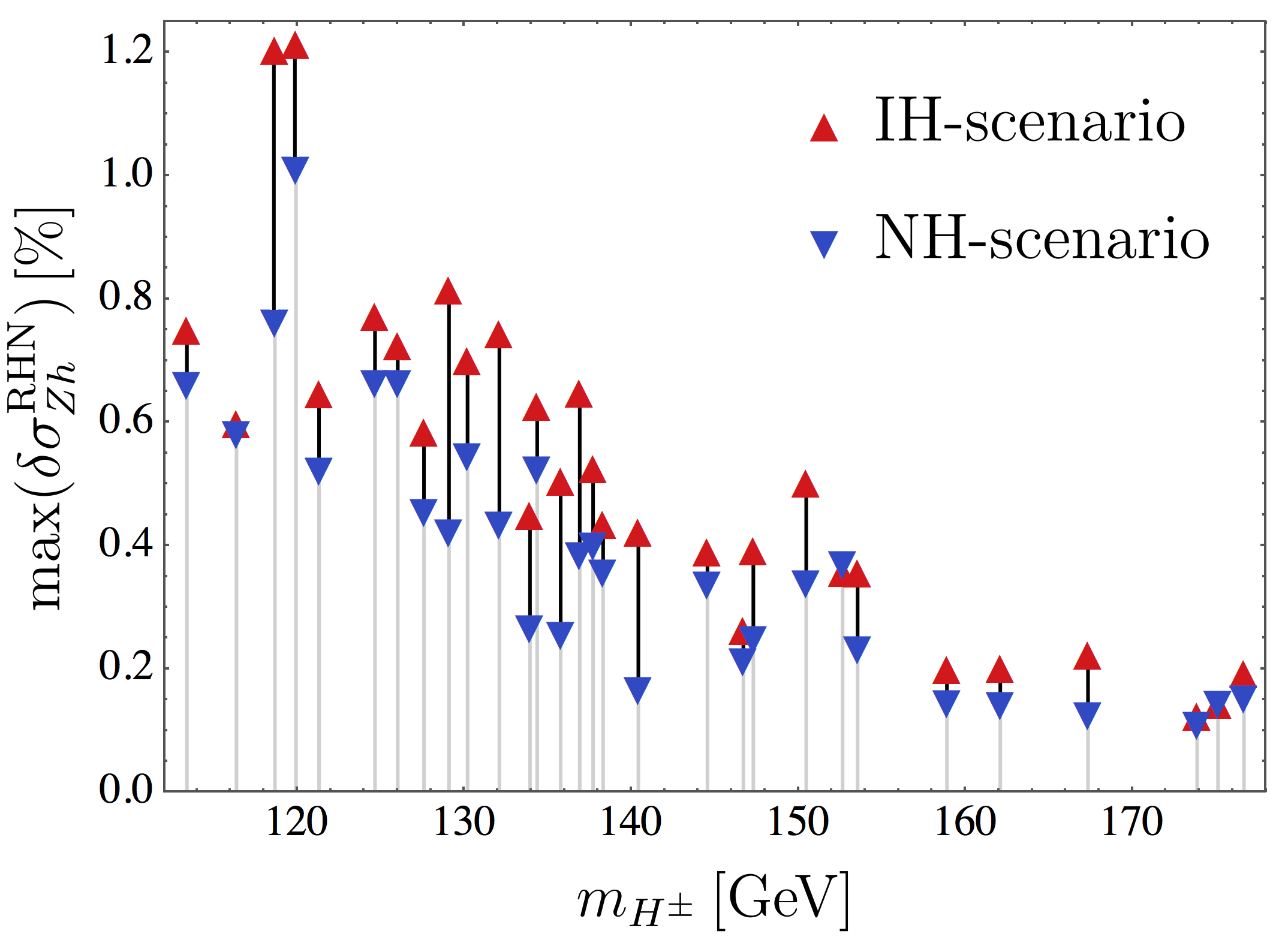}
\caption{\label{fig:dSigma_NOvsIO} \small Point with the largest $\delta \sigma^{\rm RHN}_{Zh}$ for NH-scenario (black points) and IH-scenario (red points). The blue and red points along a vertical line share the same values of $\lambda_2, \lambda_3, m_{H^0}, m_{A^0}$ and $m_{H^{\pm}}$. }
\label{fig:NOIO}
\end{figure}

\begin{figure*}
    \centering
  \includegraphics[width=0.9\textwidth]{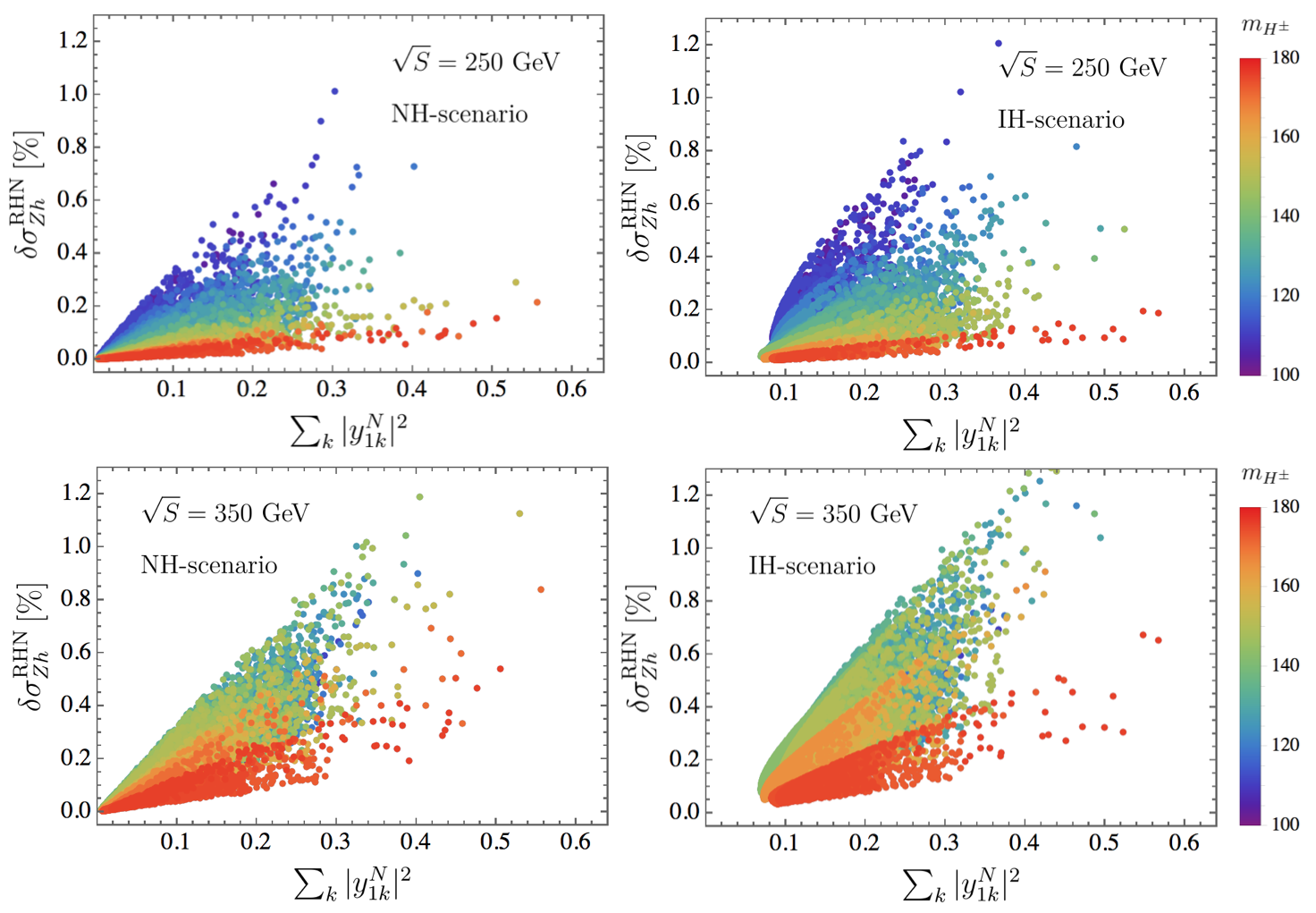}
  \caption{\label{fig:dSigma_vs_yuk} \small Difference between the Scotogenic and the IDM cross-sections (in percentage of the SM value) as a function of the neutrino Yukawa couplings. This is shown in the case of the NH-scenario (left panels) and the IH-scenario (right panels) and for two representative values of the center of mass energy: $\sqrt{S}=250$~GeV (upper panels) and $\sqrt{S}=350$~GeV (bottom panels). }
\end{figure*}
\begin{figure*}
    \centering
  \includegraphics[width=0.9\textwidth]{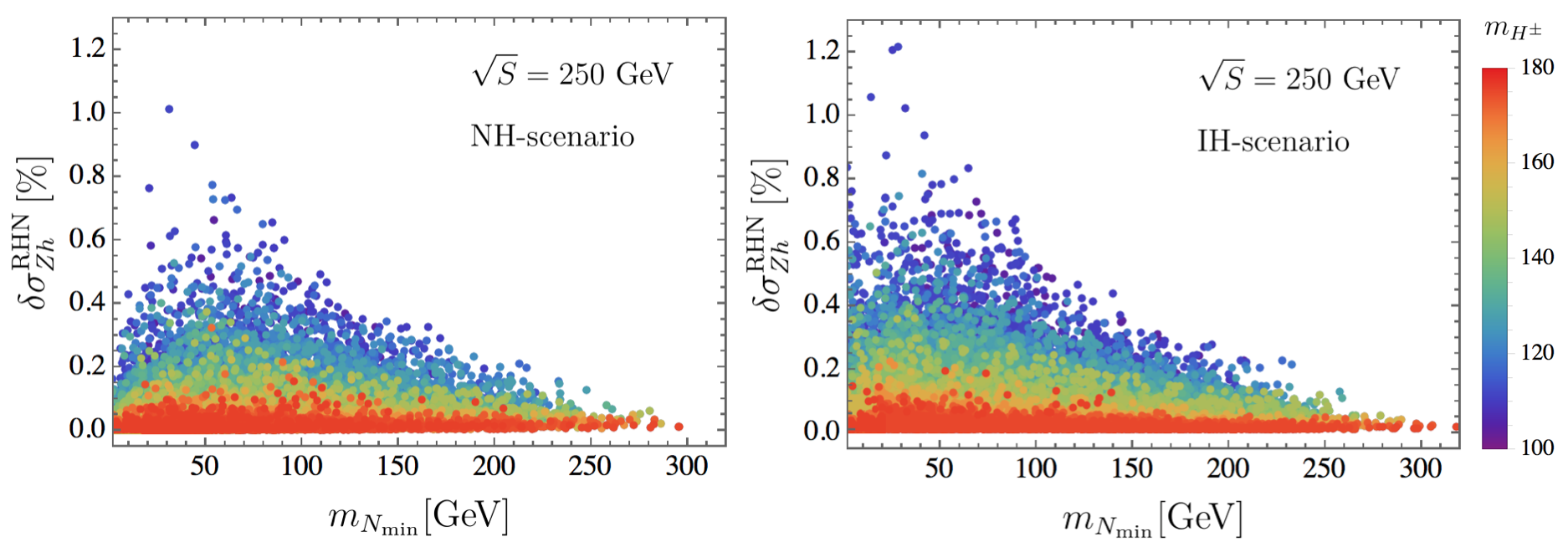}
  \caption{\label{fig:dSigma_vs_mN}
 \small Difference between the Scotogenic and the IDM cross-sections (in percentage of the SM value) as a function of the lightest RH-neutrino mass at $\sqrt{S}=250$GeV. This is shown in the case of the NH-scenario (left panel) and the IH-scenario (right panel). }
\end{figure*}

Our survey of the impact of the Scotogenic model on $e^- e^+\rightarrow Z h$ has shown, in almost all cases, to dampen the negative contribution of the IDM compared to the SM. At the same time, large values of IDM corrections are a prerequisite for large values of the corresponding Scotogenic ones. For this reason, we focus our analysis \ref{tab:scan0} on those points where, at collider energy of $\sqrt{S} = 250$ GeV,  $\sigma^{\rm IDM}_{Zh}$ accounts for more than $2\%$ of the SM value. A further restriction over $|\lambda_3| > 0.5$ still allows us to cover all the regions of masses in Table \ref{tab:scan0}. On these IDM outposts, we test the introduction of $Y_N$ and $m_N$ complying with neutrino phenomenology and we explore the patterns of inverted and normal hierarchy. 
To measure the relevance of the RHNs corrections on top of the IDM background, we make use of the quantity 
\bea
&\delta \sigma^{\rm RHN}_{Zh}  = \frac{\sigma^{\rm Scoto}_{Zh}-\sigma^{\rm IDM}_{Zh} }{\sigma^{\rm SM}_{Zh}} \, , 
\eea
with $\sigma^{\rm Model}_{Zh} = \sigma^{\rm Model}_{Zh} (S) $ the unpolarized $e^- e^+\rightarrow Z h$ cross-section computed at NLO within a specific Model. Unless explicitly stated, the reference value $\sqrt{S} = 250$ GeV will be assumed.

In Fig.~\ref{fig:dSigma_NOvsIO} we display the maximal difference $\delta \sigma^{\rm RHN}_{Zh} $ obtained in the NH (blue points) and IH (red points) scenarios as a function of the charged scalar $m_H^\pm$ for $\sqrt{S}=250$ GeV. The blue and red points along the vertical line share the same IDM input values of $\lambda_2$, $\lambda_3$, $m_{H^0}$, $m_{A^0}$ and $m_{H^\pm}$. 
The quantity $\delta \sigma^{\rm RHN}_{Zh} $ exhibits a strong dependence on the value of the $m_H^\pm$. In fact, highlighting the decoupling effect of large masses of $m_{H^{\pm}}$, as expected the largest differences are reached for the lightest charged scalar. We find that in order to obtain a $\delta \sigma^{\rm RHN}_{Zh}\gtrsim (0.2-0.4)\%$, a light charged scalar mass below $m_{H^\pm}\lesssim (170-150)$~GeV is required.
Moreover, in the range \ref{tab:scan0}, we see that it is the IH scenario that can accommodate the stronger signatures for most of the considered IDM benchmarks. This is linked to the larger volume of points we found, compared to the NH case, surviving the tight phenomenological bounds. In particular, the maximal values are reached at $m_{H^\pm}\simeq 120 $ GeV: in the NH-scenario it amounts to $\delta \sigma^{\rm RHN}_{Zh} \simeq 1.01 \% $ while in the IH scenario, we reach up to $\delta \sigma^{\rm RHN}_{Zh} \simeq 1.22 \% $.

In Fig.~\ref{fig:dSigma_vs_yuk} and \ref{fig:dSigma_vs_mN} we show again $\delta \sigma^{\rm RHN}_{Zh}$ as a function of the involved neutrino Yukawa couplings $y^N_{1k}$ $(k=1,2,3)$ and the lightest RH neutrino mass $m_{N_{\min}}\equiv\min(m_{N_1},m_{N_2},m_{N_3})$, respectively.
As expected, larger differences are reached with bigger values of the Yukawa involved in the process, as well as lighter RHN masses. Specifically, we deduce that achieving a $\delta \sigma^{\rm RHN}_{Zh}\gtrsim 0.2\% $ demands $\sum_k |y^N_{1k}|^2\gtrsim 0.056$ and $m_{N_{\rm min}}\lesssim 215$ GeV, in the case of NH, and $\sum_k |y^N_{1k}|^2\gtrsim 0.085$ and $m_{N_{\rm min}}\lesssim 233$ GeV, in the case of IH.
Within the range explored, in particular considering the narrow interval of the splitting among neutral scalars in \ref{tab:scan0}, we find a lower cutoff for the values of the Yukawa matrix generated in the IH scenario. This can be tracked to the Casas-Ibarra parametrization \ref{eq:CasasIbarraParam} and the link, in the IH case, between the $y^N_{1k}$ and the larger values of $U^{\nu}_{1k}$ and $m_{\nu}$ in Eq.~\ref{eq:CasasIbarraParam}.

\subsection{Numerical Analysis: selected benchmark points}
\label{sec:Benchmarks}

The experimental identification of a particular model requires the support of a multitude of observables beyond the mere match of a single cross-section output. It is therefore important to profile the correlations that exist between different candidate experimental signatures. The Scotogenic model has a prominent role in this, connecting the existence of heavier new states to the properties of light SM neutrinos. To illustrate these correlations, we select six benchmark points (BPs) in the region of the parameter space most likely to be (dis)proven by the combination of the measurements of LFV at future experiments and $e^+ e^-\rightarrow Zh$ at future colliders. In light of what we discussed in the previous section, this demands: (i) a light-charged scalar, (ii) a sizable value $\lambda_3$, (iii) a tiny mass splitting between the two neutral scalars of the order of the eV and (iv) at least one light RHN. 

The input values for the selected BPs are shown in Table~\ref{tab:BNCin} where we choose $m_{H^\pm}\simeq (119.9-152.6)$~GeV, $\lambda_3\simeq -(0.88-0.95)$, $m_{H^0}-m_{A^0}\simeq(41.7-78.8)$~eV, which translates into neutrino Yukawa entries of the order $|y^{N}_{ik}|\simeq (0.01-0.6)$, and $m_{N_{\rm min}}\simeq (28-109)$~GeV. Among these points, the couples $(\text{BP1},\text{BP4})$, $(\text{BP2},\text{BP5})$ and $(\text{BP3},\text{BP6})$ share the same IDM input values but correspond to different neutrino scenarios: $(\text{NH},\text{IH})$. 

In Table~\ref{tab:BNC} we show the total higgs-stralung at NLO for each one of the proposed BPs in the Scotogenic and IDM models at two representative values of the center of mass energy $\sqrt{S}= 250, 350$ GeV. Given the SM in Eq.~(\ref{eq:sigmaSM_250_350}), we have that $\sigma^{\rm Scoto}_{Zh}/\sigma^{\rm SM}_{Zh} -1 \simeq -(2.2-3.4)$~[\%] and therefore within the FCC-ee expected sensitivity, while $\delta \sigma^{\rm RHN}_{Zh} \simeq (0.25-1.2)$~[\%] and $\delta \sigma^{\rm RHN}_{Zh} \simeq (0.21-1.0)$~[\%] in the NH and IH scenarios, respectively. 
We provide also the corresponding deviations of the decay width of $ Z \rightarrow l^+l^-$ and $ Z \rightarrow inv.$.
In Table~\ref{tab:BNClfv} we list the values of the LFV processes that correspond to the selected BPs. Contrasting with the future limits in Table~\ref{tab:LFV}, we see that the predictions for $l_i \rightarrow l_j \gamma$ and $\mu \rightarrow 3 e$ lie within the expected future sensitivity for the entire set of proposed points. On the other hand $\tau \rightarrow 3 e(\mu)$ are within $(0.3-1.9)\times BR(\tau \rightarrow 3 e)^{\texttt{BelleII}}$ and $(0.02-0.34)\times BR(\tau \rightarrow 3 \mu)^{\texttt{BelleII}}$ therefore out of the experimental range explored by future experiments with the only exception of BP5 for which $\tau \rightarrow 3e$ is accessible.

\begin{figure}[t]
\includegraphics[scale=0.5]{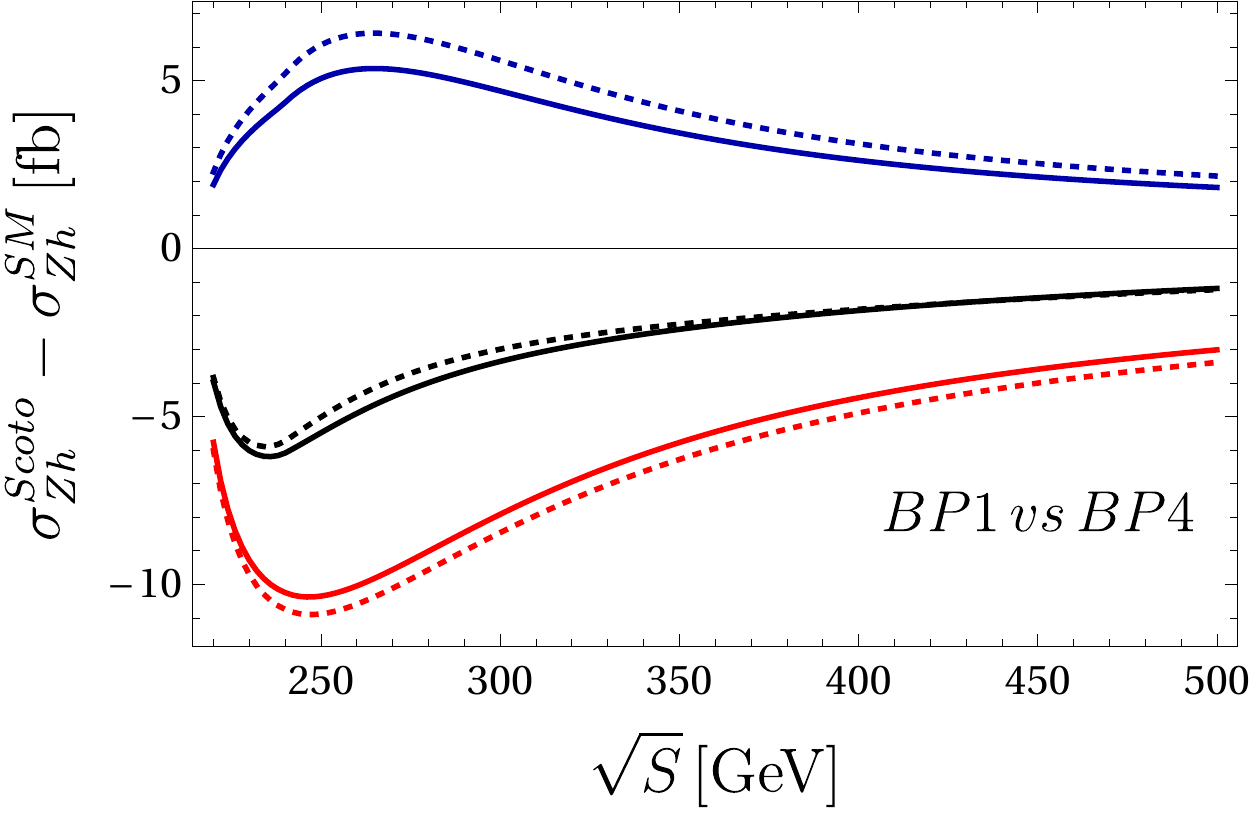}
\includegraphics[scale=0.5]{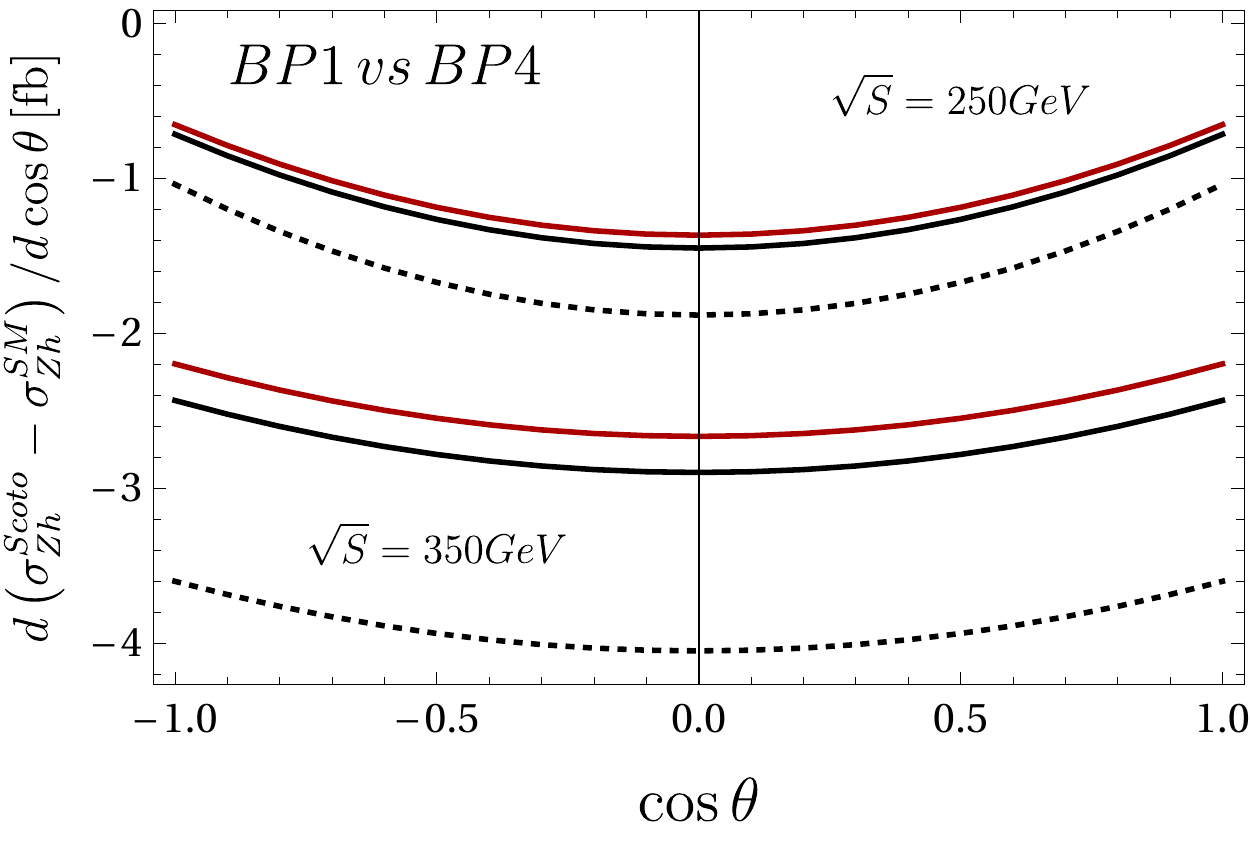}
\caption{\small Top panel: For points BP1 (continuous lines) and BP4 (dashed lines). The full $ \sigma^{\rm Scoto}_{Zh} - \sigma^{\rm SM}_{Zh}$ (in black) and the separate contribution on it of boxes (in blu) and triangles (in red). Bottom panel: For points BP1 (black lines) and BP4 (red lines). The angular distributions at 2 different collider energies. Dotted lines are the corresponding distributions for the IDM case, setting $y^N = 0$.}
\label{fig:sigma_vs_S_BP1BP4}
\end{figure}

\begin{figure}[t]
\includegraphics[scale=0.5]{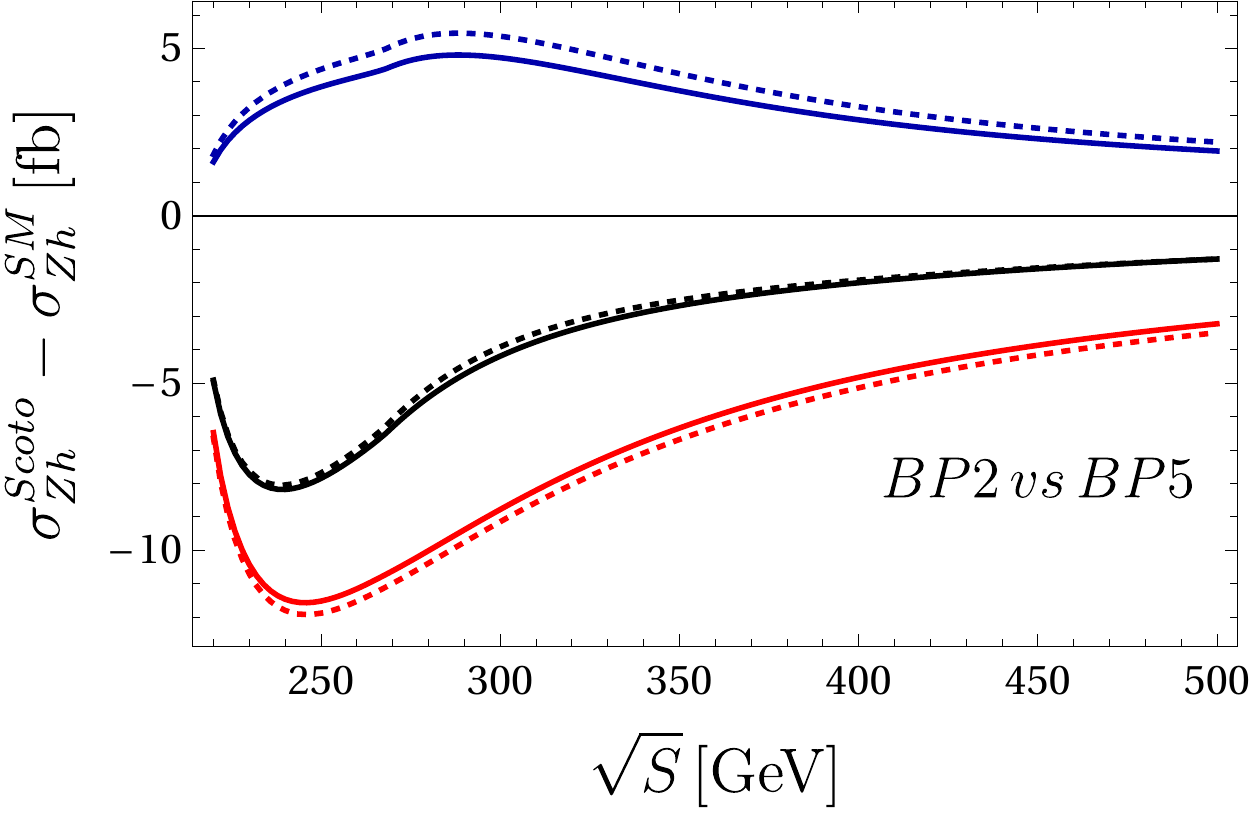}
\includegraphics[scale=0.5]{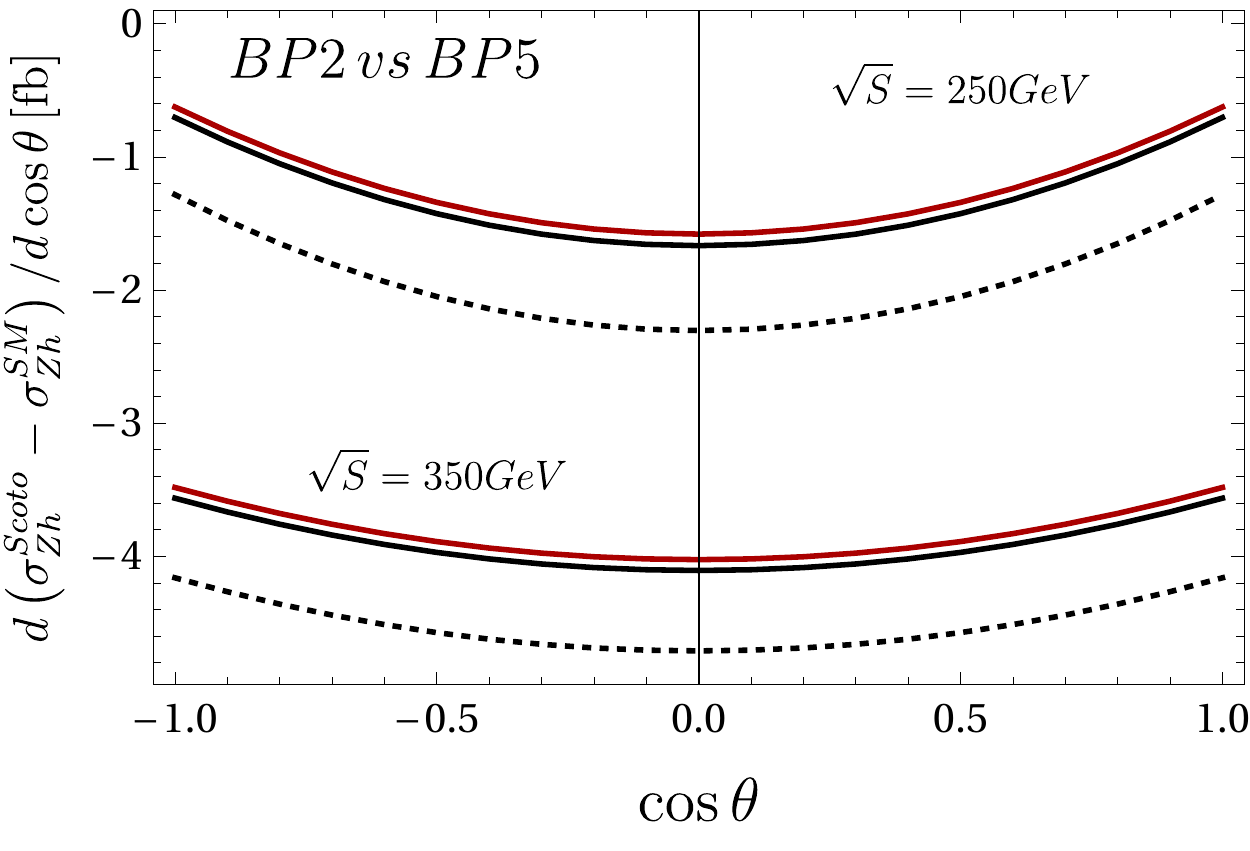}
\caption{\small For points BP2 and BP5. Same captions of Fig.~\ref{fig:sigma_vs_S_BP1BP4} with BP1 $\rightarrow$ BP2 and BP4 $\rightarrow$ BP5.}
\label{fig:sigma_vs_S_BP2BP5}
\end{figure}

\begin{figure}[t]
\includegraphics[scale=0.5]{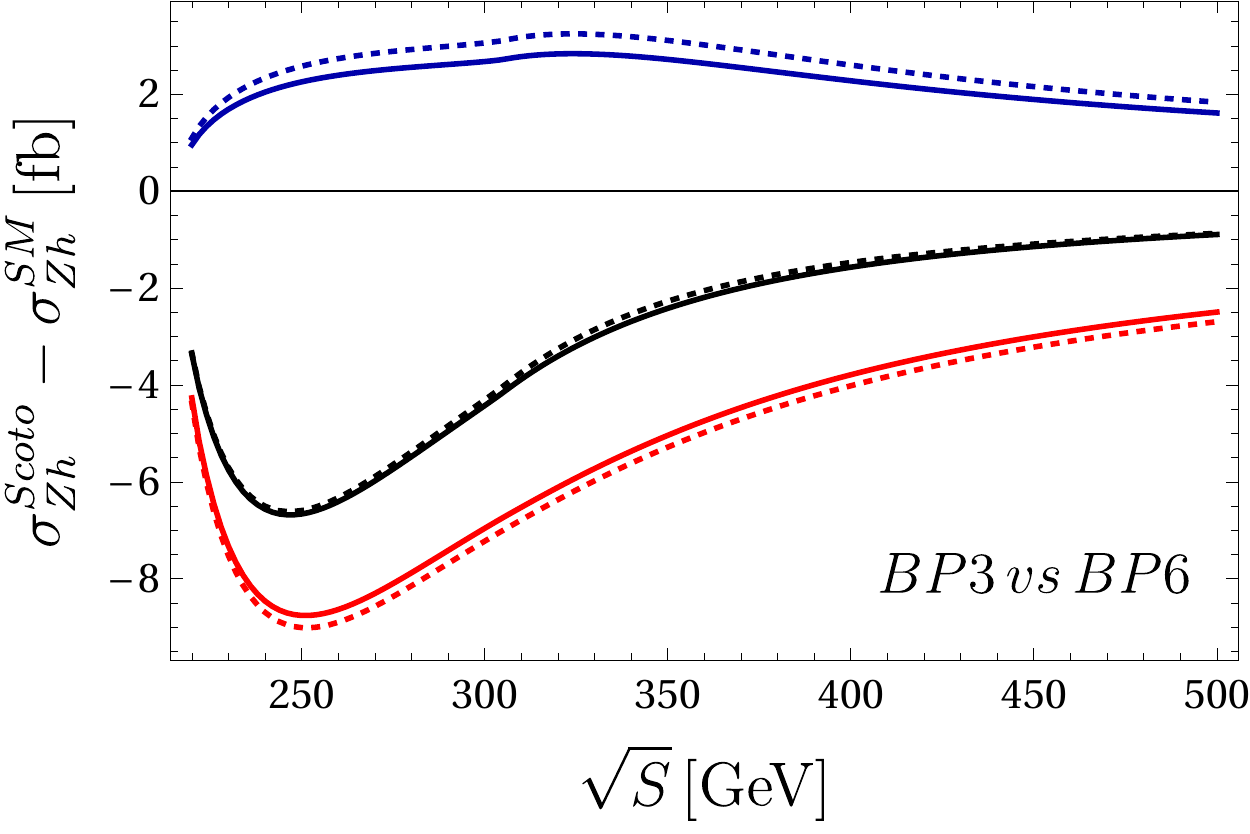}
\includegraphics[scale=0.5]{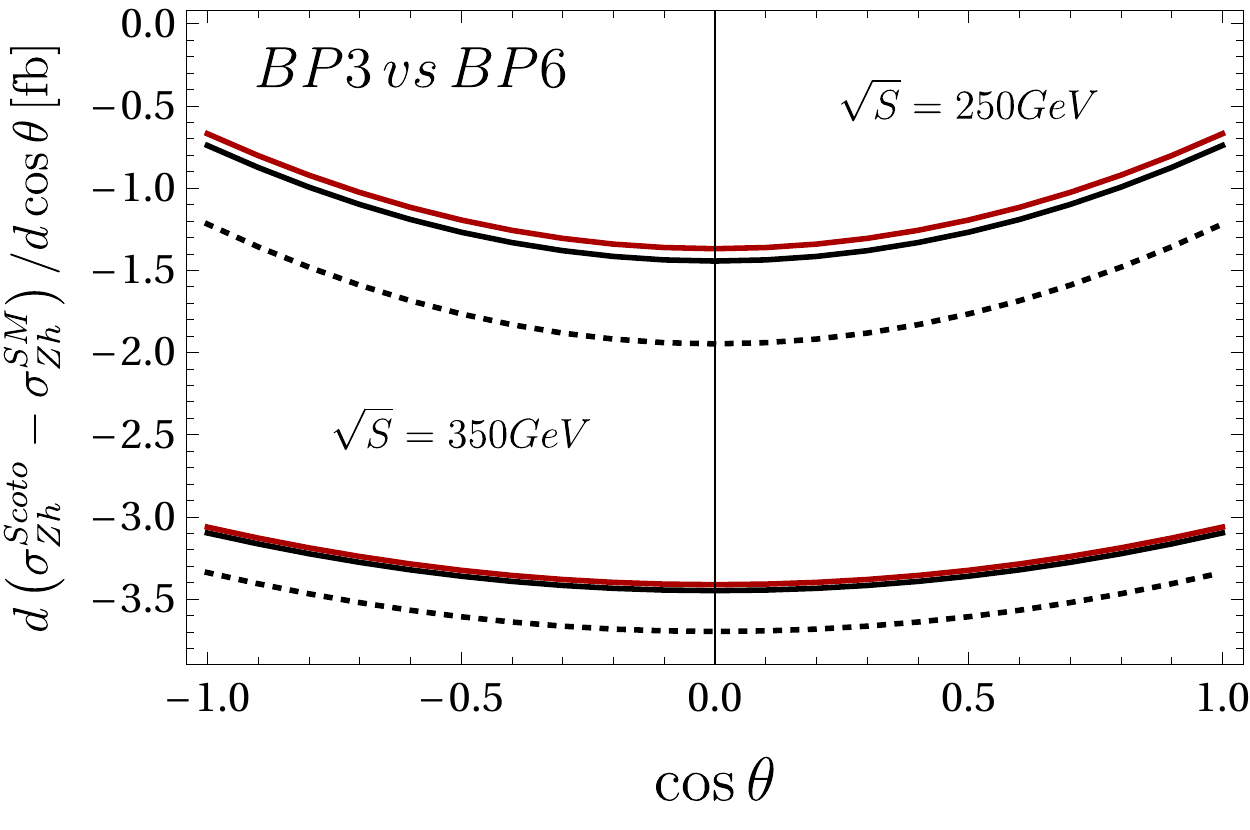}
\caption{For points BP3 and BP6. Same captions of Fig.~\ref{fig:sigma_vs_S_BP1BP4} with BP1 $\rightarrow$ BP3 and BP4 $\rightarrow$ BP6.}
\label{fig:sigma_vs_S_BP3BP6}
\end{figure}

Eventually, for these points, in Figures~\ref{fig:sigma_vs_S_BP1BP4}- \ref{fig:sigma_vs_S_BP3BP6} we study the full $\sigma^{\rm Scoto}_{Zh}-\sigma^{\rm SM}_{Zh}$ as a function of $\sqrt{S}$ (top panel) and the angular dependence of the corresponding differential quantity.

\begin{table*}[t]
	\centering
 	\resizebox{1\textwidth}{!}{
    \setlength{\tabcolsep}{0pt}
     \renewcommand{\arraystretch}{1.1}
	\begin{tabular}{|c|c c c c c c|}
    \hline
    \bf Input   & \bf BP1 & \bf BP2 & \bf BP3 & \bf BP4  &\bf BP5  &\bf BP6 \\[1.pt]
    \bf values  &  NH &  NH &  NH &  IH &  IH &   IH\\
    \hline
    $m_{H^{\pm}} [\text{GeV}]$             & 119.923    & 134.361  & 152.644  & 119.923 & 134.361  & 152.644 \\
    $m_{H_{0}} [\text{GeV}]$               & 96.606     & 98.174   & 112.314  & 96.606  & 98.174   & 112.314 \\
    $m_{H_{0}}$- $m_{A_{0}} [\text{eV}]$   & 78.803     & 51.414   & 41.727   & 78.803  & 51.415   & 41.727  \\
    $\lambda_2$                            & 3.152      & 3.530    & 3.175    & 3.152   & 3.525    & 3.175   \\
    $\lambda_3$                            & -0.878     & -0.951   & -0.884   & -0.878  & -0.951   & -0.884  \\
    $m_{N_1} [\text{GeV}]$                 & 39.991     & 42.387   & 225.781  & 27.937  & 243.846  & 223.871 \\
    $m_{N_2} [\text{GeV}]$                 & 31.172     & 190.267  & 100.971  & 156.549 & 31.992   & 109.049 \\
    $m_{N_3} [\text{GeV}]$                 & 53.728     & 33.420   & 300.580  & 29.658  & 41.194   & 147.786 \\
    $Y_N/0.1$                              & $\scriptsize\begin{array}{c}\{\{1.97,5.03,1.03\},\\\{-3.30,0.546,3.82\},\\\{3.39,-2.41,3.13\}\}\end{array}$       & $\scriptsize\begin{array}{c}\{\{2.05,-0.230,-5.31\},\\\{-0.860,-4.30,-0.244\},\\\{-5.12,0.114,-2.51\}\}\end{array}$     & $\scriptsize\begin{array}{c}\{\{2.00,5.44,1.90\},\\\{-4.26,0.281,4.11\},\\\{3.25,-2.97,4.05\}\}\end{array}$     & $\scriptsize\begin{array}{c}\{\{5.92,-0.291,-0.750\},\\\{0.694,0.836,5.38\},\\\{-0.435,-3.78,0.781\}\}\end{array}$ & $\scriptsize\begin{array}{c}\{\{0.375,4.93,3.53\},\\\{-3.76,-0.954,1.52\},\\\{1.96,-3.53,3.46\}\}\end{array}$ & $\scriptsize\begin{array}{c}\{\{0.926,5.79,2.61\},\\\{-4.99,-0.700,3.12\},\\\{3.29,-2.79,4.29\}\}\end{array}$\\
    \hline
\end{tabular}}
\caption{\label{tab:BNCin}\small Input values for the selected benchmark points consistent with $Z$ and $W$ decays in Table~\ref{tab:contsraints}, neutrino oscillation observables in Table~\ref{tab:neutrinoData}, LFV constraints in Table~\ref{tab:LFV}.}
\end{table*}
\begin{table*}[t]
	\centering
 \resizebox{0.8\textwidth}{!}{
     \setlength{\tabcolsep}{10pt}
     \renewcommand{\arraystretch}{1.3}
	\begin{tabular}{|c| c c c c c c|}
    \hline
    \bf  \multirow{2}{*}{Cross-section} & \bf BP1 & \bf BP2 & \bf BP3 & \bf BP4 & \bf BP5 & \bf BP6 \\[1.pt]
       &  NH &  NH &  IH &  IH &  IH &  IH\\
    \hline
    $\sigma^{\rm Scoto}_{Zh}$(250 GeV) [fb]  & 223.261   & 220.897   & 222.085   & 223.729   & 221.060   & 222.157   \\
    $\sigma^{\rm IDM}_{Zh}$(250 GeV) [fb]    & 220.946   & 219.693   & 221.593   & 220.946   & 219.693   & 221.593 \\
    $\delta\sigma^{\rm RHN}_{Zh}$(250 GeV)   & 1.0122\%  & 0.5262\%  & 0.2150\%  & 1.2168\%  & 0.5973\%  & 0.2464\%  \\
    \hline
    $\sigma^{\rm Scoto}_{Zh}$(350 GeV) [fb]  & 120.986   & 120.702   & 120.973   & 121.138   & 120.868   & 121.121   \\
    $\sigma^{\rm IDM}_{Zh}$(350 GeV) [fb]    & 120.191   & 119.464   & 119.983   & 120.191   & 119.464   & 119.983 \\
    $\delta\sigma^{\rm RHN}_{Zh}$(350 GeV)   & 0.6442\%  & 1.0016\%  & 0.8024\%  & 0.7678\%  & 1.1377\%  & 0.9222\%  \\
    \hline
    $\delta\Gamma(Z\rightarrow ee)^{\rm RHN}$  [MeV]         & 0.04388   & 0.08528   & 0.09608   & 0.04026   & 0.08453   & 0.09552   \\ 
    $\delta\Gamma(Z\rightarrow \mu\mu)^{\rm RHN}$   [MeV]    & 0.04502   & 0.08861   & 0.09738   & 0.04390   & 0.08892   & 0.09693   \\ 
    $\delta\Gamma(Z\rightarrow \tau\tau)^{\rm RHN}$  [MeV]   & 0.04459   & 0.08518   & 0.09687   & 0.04803   & 0.3916    & 0.08615    \\ 
    $\delta\Gamma(Z\rightarrow inv.)^{\rm RHN}$    [MeV]     & 0.1678    & 0.3724    & 0.4419    & 0.1734    & 0.3754    & 0.4318    \\ 
    \hline
\end{tabular}}
\caption{\label{tab:BNC} \small One loop total cross-section $e^+e^- \rightarrow Zh$ in the Scotogenic and IDM models for the BPs in Table~\ref{tab:BNCin}. The values are given at two different centers of mass energy $\sqrt{S}=250, 350$ GeV. Below we display the corresponding deviations of the decay width of $Z\rightarrow l^+l^-$ and $Z\rightarrow inv.$ .}
\vspace{5mm}
\resizebox{0.9\textwidth}{!}{
     \setlength{\tabcolsep}{10pt}
     \renewcommand{\arraystretch}{1.3}
	\begin{tabular}{|c| c c c c c c|}
    \hline
    \bf  \multirow{2}{*}{LFV process} & \bf BP1 & \bf BP2 & \bf BP3 & \bf BP4 & \bf BP5 & \bf BP6 \\[1.pt]
       &  NH &  NH &  IH &  IH &  IH &  IH\\
    \hline
    BR($\mu\rightarrow e \gamma$)          & $3.24\times 10^{-13}$ & $1.02\times 10^{-13}$ & $2.28\times 10^{-13}$ & $2.43\times 10^{-13}$ & $6.22\times 10^{-14}$ & $1.05\times 10^{-13}$ \\
    BR$(\tau\rightarrow e\gamma)$          & $1.38\times 10^{-8}$  & $1.09\times 10^{-8}$  & $3.06\times 10^{-8}$  & $1.16\times 10^{-8}$  & $3.10\times 10^{-8}$  & $8.39\times 10^{-9}$  \\
    BR$(\tau\rightarrow\mu\gamma)$         & $3.60\times 10^{-9}$  & $2.09\times 10^{-8}$  & $1.02\times 10^{-9}$  & $9.90\times 10^{-9}$  & $3.78\times 10^{-8}$  & $3.89\times 10^{-9}$  \\
    \hline
    BR($\mu\rightarrow 3 e$)               & $3.52\times 10^{-13}$ & $1.63\times 10^{-13}$ & $5.71\times 10^{-14}$ & $5.37\times 10^{-13}$ & $4.13\times 10^{-13}$ & $1.00\times 10^{-13}$ \\
    BR$(\tau \rightarrow 3 e)$             & $3.32\times10^{-10}$  & $2.73\times 10^{-10}$ & $4.96\times 10^{-10}$ & $3.11\times 10^{-10}$ & $8.80\times 10^{-10}$ & $1.44\times 10^{-10}$ \\
    BR$(\tau\rightarrow 3\mu)$             & $5.29\times10^{-11}$  & $5.43\times 10^{-11}$ & $1.02\times 10^{-11}$ & $1.55\times 10^{-10}$ & $1.54\times 10^{-10}$ & $3.33\times 10^{-11}$ \\
    \hline
    BR$(Z\rightarrow \mu e)$               & $1.80\times 10^{-15}$ & $3.71\times 10^{-15}$ & $1.07\times 10^{-16}$ & $1.01\times 10^{-15}$ & $4.21\times 10^{-15}$ & $4.89\times 10^{-16}$ \\
    BR$(Z\rightarrow \tau e)$              & $1.04\times 10^{-12}$ & $8.92\times 10^{-13}$ & $2.16\times 10^{-12}$ & $9.47\times 10^{-13}$ & $2.59\times 10^{-12}$ & $6.26\times 10^{-13}$ \\
    BR$(Z\rightarrow \tau \mu)$            & $1.39\times 10^{-13}$ & $2.15\times 10^{-12}$ & $3.51\times 10^{-14}$ & $4.98\times 10^{-13}$ & $2.00\times 10^{-12}$ & $1.61\times 10^{-13}$ \\
    \hline
    BR$(h\rightarrow \mu e)$               & $7.75\times 10^{-15}$ & $2.18\times 10^{-14}$ & $4.91\times 10^{-16}$ & $5.05\times 10^{-15}$ & $2.47\times 10^{-14}$ & $2.53\times 10^{-15}$ \\
    BR$(h\rightarrow \tau e)$              & $4.33\times 10^{-9}$  & $4.11\times 10^{-9}$  & $8.80\times 10^{-9}$  & $3.94\times 10^{-9}$  & $1.16\times 10^{-8}$  & $2.50\times 10^{-9}$  \\
    BR$(h\rightarrow \tau \mu)$            & $8.12\times 10^{-10}$ & $9.27\times 10^{-9}$  & $1.99\times 10^{-10}$ & $2.55\times 10^{-9}$  & $1.12\times 10^{-8}$  & $8.35\times 10^{-10}$ \\   
    \hline
\end{tabular}}
\caption{\label{tab:BNClfv} \small Values of the LFV processes that correspond to the selected BPs.}
\end{table*}

\section{Conclusions}
\label{sec:conclusions}
The presence of extra scalar states is a common ingredient for many extensions of the SM. For a large class of them
the IDM represents a limit that can effectively describe their phenomenology. 

In light of some of the SM shortcomings, such a minimal scalar extension will likely provide a portal into further states and interactions. 
Indeed, an undeniable sign of physics beyond the SM, and probably the most convincing until now, is the experimental fact that neutrinos oscillate, i.e. the SM needs to be extended to include neutrino masses and the large mixing angles that we observe.

The Scotogenic model supports the inert scalar doublet with right-handed sterile fermions, activating a mechanism for explaining the smallness of SM neutrino masses through radiative corrections.   
The new particles and the new interactions, required to comply with neutrino phenomenology, can be at the reach of forthcoming collider searches, in particular within the precise profiling expected at FCC-ee. 
It is therefore important to identify, from the IDM background, the presence of extra states uniquely linked to the Scotogenic paradigm.  

In this work, we have presented a detailed profiling of precisely those regions of the Scotogenic parameter space that can result in a clear, and distinct from the IDM, signal in Higgs-strahlung at FCC-ee. 
In order to do so we have computed the complete one-loop radiative corrections to the process. This step required the full renormalization of the parameters of the SM electroweak sector and we adopted, as customary, an OS-scheme. We have presented the full analytical formulas for the one-loop process as well as the needed counterterms. In this regard, we are also providing the community with a useful tool for further NLO investigations concerning the Scotogenic as well as sibling models sharing a similar Yukawa structure. 
We have accompanied our collider profiling demanding the radiative generation of neutrino masses and angles to fulfil the current up-to-date experimental bounds, for both normal and inverted hierarchies. This inevitably introduced sources of LFV that we have thoroughly included and targeted, so to connect a possible FCC-ee signal to the corresponding one in LFV processes.   
We have found that the inclusion of the non-universal contributions 
generated by virtual RHNs tends to interfere with the IDM, generating an overall signal in Higgs-strahlung closer to that of the SM. Still, this signal can comfortably exceed the detection threshold and, importantly, separate itself from a corresponding purely IDM scenario. 
The LFV process $\mu \rightarrow e \gamma$ presented the biggest challenge but, even accounting for all the neutrino constraints, a surmountable one. In the case study, that of 3 RHN, it appears highly unlikely that the collider signal would not present itself without a parallel one in $\mu \rightarrow e \gamma$, and we have biased the generation of the dataset by targeting a scale of BR$(\mu \rightarrow e \gamma)$ of next probing. We stress that an easier route to relax the tight constraints of LFV is to consider a generation number of RHNs greater than three. Most of our analytical formulas have been provided with a generic generation number of RHN for such a purpose.  
We can also consider the Peccei-Quinn limit $\lambda_5 =0$ and decouple the Higgs-strahlung process from the explanation of neutrino phenomenology. In this case, our formulas would describe the impact of non-universal contributions from RHNs to $e^- e^+\rightarrow Z h$ which is a common feature of many BSM proposals. 

Finally, we have selected and presented benchmark points so as to highlight, in more detail, the differences between the neutrino mass hierarchies, the interplay with the LFV observables computed, and the energy and angle-dependent cross-sections. 

We leave to future work the interesting assessment of the impact of CP violations on the polarized cross-section and the related neutrino and leptonic low-energy observables. 

\vspace{3mm}
\noindent \textbf{Acknowledgement.} 
The work of C.M. was supported by the Estonian Research Council grant PRG1677.
We thank Marco Piva for useful discussions.

\section{Appendix}

\subsection{On-shell renormalization} \label{OSRenoApp}
We illustrate here the details of the renormalization procedure adopted in this paper. 
For the subset of the model parameters already included in the EW sector 
\bea
& g_1, g_w, Y_u, Y_d, Y_{e}, v_{h}, \lambda, \mu_1^2 \,, \nn 
\eea
we perform a trade for the alternative set
\bea \label{varSM}
&x \in \left[m^2_Z, m^2_W, {\rm e}, m_{u_i}, m_{d_i}, m_{l=e,\mu,\tau}, m^2_h, t  \right] \, . 
\eea
As they are, such parameters do not enter the final finite amplitude. A splitting in the form  $x \rightarrow x_{r} + \delta x_{\rm UV}$ is instead forced to absorb in $\delta x_{\rm UV}$ the (UV) regulator dependence, and leave the amplitude as a function of the renormalized, finite coupling $x_r$. Applying this splitting to the set \ref{varSM}   
\bea \label{CoRen}
& m^2_Z \rightarrow m^2_Z  + \delta m^2_Z, \,\, m^2_W \rightarrow m^2_W  + \delta m^2_W,  \\
& m_f \rightarrow m_f  + \delta m_f, \quad (f = u,d,l)  \\
& m^2_h \rightarrow m^2_h  + \delta m^2_h, \,\, t \rightarrow t  + \delta t \, ,  \\
& {\rm e} \rightarrow {\rm e}\left(1 + \delta {\rm e}\right) \,,
\eea
we generate the counterterms needed to define a finite S-matrix element for $e^+ e^- \rightarrow Z h$.  
Moreover, we work at all the stages with finite Green functions, by introducing further counterterms via the field rescaling \footnote{For the model and the approximation involved in our study, we can work with a flavor diagonal rescaling and ignore the renormalization of the full ghost and QCD sector.}
\bea \label{FiRen}
& W_{\mu} \rightarrow \left( 1 + \frac{\delta_W}{2}\right)W_{\mu}  ,  \\
& Z_{\mu} \rightarrow \left( 1 + \frac{\delta_Z}{2}\right)Z_{\mu} + \frac{1}{2}\delta_{ZA} A_{\mu},  \\ 
& A_{\mu} \rightarrow\left( 1 + \frac{\delta_A}{2}\right)A_{\mu} + \frac{1}{2}\delta_{AZ} Z_{\mu},  \\ 
& \Phi \rightarrow \left( 1 + \frac{\delta_h}{2}\right)\Phi , \quad (\Phi = h, G^0, G^{\pm})  \\
& \psi \rightarrow \left( 1 + \delta_{\psi}\right)\psi, \quad (\psi = e_L, e_R, \nu_L) \, .
\eea
The \emph{On-Shell} nature of the scheme is granted by demanding that the counterterms secure, at all orders, that $m_Z,m_W,m_h,m_f$ can indeed be identified with the poles of the two-point functions of the EW gauge bosons, the Higgs field and the SM Dirac fermions. Then, the field renormalizations of \ref{FiRen} are constrained by fixing the residue of the corresponding amplitude over such poles. 
The QED coupling constant $e_{QED}$ is fixed by the vertex $e \gamma \rightarrow e $ in the Thomson limit. QED gauge symmetry enforces the following connection to two-point function renormalization constants
\bea
\delta {\rm e} = - \frac{\delta_{A}}{2} - \delta_{ZA} \frac{s_W}{c_W} \,.
\eea
Lastly, the parameter $t$ is fixed through the relation $\mu_1^2 = \frac{m_h^2}{2} + \frac{t}{v_{h}}$. The OS relation $m^2_h = 2 \mu_1^2$ is enforced by demanding $t=0$, and is kept at all order by absorbing in $\delta t$ the momentum-independent Higgs tadpole diagrams. 

\subsection{Counterterms in the IDM}
\label{appIDMCT}

Here we provide the explicit form of the counterterms in the pure IDM scenario. At one-loop this is equivalent in the Scotogenic case since the two-point functions of the EW bosons do not receive corrections from the right-handed neutrinos. These are computed by imposing 
\bea
\delta m^2_V=\widetilde{\Sigma}^V_T(m^2_V),\quad \text{and}\quad \delta_V=\frac{\partial \widetilde{\Sigma}^V_T}{\partial k^2}\bigg|_{k^2=m^2_V},
\eea
where the \emph{tilde} selects the real part of the loop integrals. We find
\bea
\delta m^2_Z &= \frac{\alpha}{16\pi c^2_Ws^2_W}\bigg[ 2(c^2_W-s^2_W)^2A_0[m^2_{H^\pm}]\\
&+A_0[m^2_{H^0}] +A_0[m^2_{A^0}]-4 \widetilde{B}_{00}[m^2_Z,m^2_{A^0},m^2_{H^0}]\\
&-4(c^2_W-s^2_W)^2\widetilde{B}_{00}[m^2_Z,m^2_{H^\pm},m^2_{H^\pm}]\bigg],
\eea
\bea
\delta_Z &= \frac{\alpha}{4\pi c^2_W s^2_W}\bigg(\widetilde{B}'_{00}[m^2_Z,m^2_{A^0},m^2_{H^0}]\\
&+(c^2_W-s^2_W)\widetilde{B}'_{00}[m^2_Z,m^2_{H^\pm},m^2_{H^\pm}]\bigg),
\eea
\bea
\delta m^2_W &=\frac{\alpha}{16\pi s^2_W}\bigg(2A_0[m^2_{H^\pm}]+A_0[m^2_{H^0}]+A_0[m^2_{A^0}]\\
&+4 \widetilde{B}_{00}[m^2_W,m^2_{H^\pm},m^2_{H^0}]\\
&+4 \widetilde{B}_{00}[m^2_W,m^2_{A^0},m^2_{H^\pm}]\bigg),
\eea
\bea
\delta_W=\frac{\alpha}{4\pi s^2_W}\bigg( \widetilde{B}'_{00}[m^2_W,m^2_{H^\pm},m^2_{H^0}]\\
+\widetilde{B}'_{00}[m^2_W,m^2_{A^0},m^2_{H^\pm}]\bigg),
\eea
\bea
\delta_{A} &= \frac{\alpha}{\pi} \widetilde{B}'_{00}[0,m^2_{H^{\pm}},m^2_{H^{\pm}}],
\eea
\bea
\delta_{A Z} =-\delta_{ZA} &= \frac{\alpha(c^2_W-s^2_W) }{2\pi c_Ws_W m_Z}\bigg( A_0[m^2_{H^\pm}]\\
&-2\widetilde{B}_{00}[m^2_Z,m^2_{H^\pm},m^2_{H^\pm}]\bigg),
\eea
\bea
\delta m^2_h &= \frac{1}{32\pi^2}\bigg( 2\lambda_3 A_0[m^2_{H^\pm}]+(\lambda_3+\lambda_4+\lambda_5) A_0[m^2_{H^0}]\\
&+(\lambda_3+\lambda_4-\lambda_5) A_0[m^2_{A^0}]\bigg)\\
&+\frac{m^2_Ws^2_W}{32\pi^3\alpha}\bigg(2\lambda^2_3 \widetilde{B}_0[m^2_h,m^2_{H^\pm},m^2_{H^\pm}]\\
&+(\lambda_3+\lambda_4+\lambda_5)^2 \widetilde{B}_0[m^2_h,m^2_{H^0},m^2_{H^0}]\\
&+(\lambda_3+\lambda_4-\lambda_5)^2 \widetilde{B}_0[m^2_h,m^2_{A^0},m^2_{A^0}]\bigg),
\eea
\bea
\delta_{h} &= -\frac{m^2_W s^2_W}{32\pi^3\alpha }\bigg( 2\lambda^2_3 \widetilde{B}'_0[m^2_h,m^2_{H^\pm},m^2_{H^\pm}] \\
&+(\lambda_3+\lambda_4 +\lambda_5)^2 \widetilde{B}'_0[m^2_h,m^2_{H^0},m^2_{H^0}]\\
&+(\lambda_3+\lambda_4 -\lambda_5)^2 \widetilde{B}'_0[m^2_h,m^2_{A^0},m^2_{A^0}] \bigg),
\eea
\bea
\delta t = -\frac{m_W s_W}{16{\rm e}\pi^2}\bigg[& 2\lambda_3A_0[m^2_{H^\pm}]+(\lambda_3+\lambda_4+\lambda_5)A_0[m^2_{H^0}]\\
&+(\lambda_3+\lambda_4-\lambda_5)A_0[m^2_{A^0}] \bigg],
\eea
\bea
\delta_{G^0} &= -\frac{m^2_Ws^2_W}{16 \pi^3 \alpha}\lambda^2_5 \widetilde{B}'_0[m^2_Z,m^2_{A^0},m^2_{H^0}],
\eea
\bea
\delta_{G^\pm} &=-\frac{m^2_W s^2_W}{64\pi^3\alpha}\bigg[ (\lambda_4+\lambda_5)^2 \widetilde{B}'_{0}[m^2_W,m^2_{H^\pm},m^2_{H^0}]\\
&+(\lambda_4-\lambda_5)^2 \widetilde{B}'_{0}[m^2_W,m^2_{A^0},m^2_{H^\pm}]\bigg].
\eea
\subsection{Counterterms at order $|y^N_{ik}|^2$} \label{appCT}
We present here the explicit formulas for the subset of counterterms in \ref{CoRen} and \ref{FiRen} receiving a non-universal correction of order $|y^N_{ik}|^2$. The notation of the scalar loop integrals is taken from \cite{Hahn:2000jm,Hahn:2000kx,Hahn:2016ebn}. 
We find
\bea
 \delta m_{i = e,\mu, \tau} & = \frac{1}{64 \pi^2 m_i} \sum_{k}  |y^N_{ik}|^2 \bigg[   A_0[m^2_{H^\pm}]-A_0[m_{N_k}^2]  \nn \\
& + \big( m_i^2 - m^2_{H^\pm} +  m_{N_k}^2 \big) + \tilde{B}_{0}[m_i^2, m^2_{H^\pm}, m_{N_k}^2] \bigg]\,,
\eea
\bea
\delta \psi_{L i = e,\mu, \tau} &= - \frac{1}{32 \pi^2} \sum_{k}  |y^N_{ik}|^2 \bigg[   \tilde{B}_0[m_i^2, m^2_{H^\pm}, m_{N_k}^2]   \nn \\
& + \big( m_i^2 - m^2_{H^\pm} +  m_{N_k}^2 \big) + \tilde{B}'_{0}[m_i^2, m^2_{H^\pm}, m_{N_k}^2] \bigg]\,,
\eea
\bea
\delta \psi_{R i = e,\mu, \tau} &= \frac{1}{32 \pi^2 m_i^2} \sum_{k}  |y^N_{ik}|^2 \bigg[   A_0[m^2_{H^\pm}]-A_0[m_{N_k}^2]  \nn \\
& + \big(- m^2_{H^\pm} +  m_{N_k}^2 \big) + \tilde{B}_{0}[m_i^2, m^2_{H^\pm}, m_{N_k}^2]  \nn \\ 
& m_i^2 \left( m_i^2 - m^2_{H^\pm} +  m_{N_k}^2 \right) \tilde{B}'_{0}[m_i^2, m^2_{H^\pm}, m_{N_k}^2]
\bigg] \,,
\eea
and similarly for SM neutrinos
\bea
& \delta \psi_{L i =\nu_{e},\nu_{\mu}, \nu_{\tau}} =  \frac{1}{32 \pi^2} \sum_{k}  |y^N_{ik}|^2  \bigg[   \tilde{B}_0[0, m^2_{H^0}, m_{N_k}^2]  \nn \\
&+\tilde{B}_1[0, m^2_{H^0}, m_{N_k}^2] + \tilde{B}_0[0, m^2_{A^0}, m_{N_k}^2] + \tilde{B}_1[0, m^2_{A^0}, m_{N_k}^2] \bigg]\,.
\eea

\subsection{RHN corrections to $Z \rightarrow l^+ l^-$ and $Z \rightarrow \nu \nu$}
\label{app:Zll_Znunu}
The flavor diagonal decays of the $Z$ admit radiative corrections and need to be renormalized in order to extract the NP contributions of the corresponding partial widths. In order to accommodate for a parameter space suitable for discovery from a precision determination of $\sigma_{Zh}$, we have assumed all the other current detection scenarios to be subdominant. Therefore we impose, for the NP-modified partial width of the $Z$ gauge boson, to be within the $2 \sigma$ experimental interval, as taken from \cite{ParticleDataGroup:2020ssz}. 
The two body decay rate for $Z(\lambda)\rightarrow f(s_+)f(s_-)$ is computed as
\bea
\label{eq:GammaZfifj}
\Gamma(Z\rightarrow f_if_i) = \frac{1}{48\pi m_Z}\left(\frac{1}{4} \sum_{\lambda,s_+, s_-} |\mathcal M\left(\lambda; s_+, s_-\right)|^2\right)
\eea
with the produced fermions being either $f=l,\nu$. At tree-level we have that the LO amplitudes can be expanded in terms of the same two matrix elements $\mathcal{M}_{L,R}$ of Eq.~(\ref{eq:HelBasis}) with coefficients given by
\bea
F^{Z\rightarrow f_if_i} _{L,R} = {\rm e} g^{Zf}_{L,R} 
\eea
where $g^{Ze}_{L,R}$ are the $Z$-boson coupling to charged leptons given in Eq.~(\ref{eq:geLR}) and $g^{Z\nu}_{L}=1/2c_W s_W$ those to neutrinos. The decay rate is readily computed through (\ref{eq:GammaZfifj}) using that $(1/4)\sum_{\lambda,s_+,s_-}|\mathcal{M}_{L,R}|^2=2m^2_Z$.
For these processes, we find that the Scotogenic states affect mainly the helicity amplitude $\mathcal{M}_{L}$ in Eq.~(\ref{eq:HelBasis}) through one-loop corrections depicted in Fig.~\ref{fig:Zll_Znunu}.
The corresponding correction to the scalar coefficient for charged leptons is
\bea
\delta F_{L}^{Z\rightarrow l_il_i} & = \frac{{\rm e}g^{Ze}_L }{8 \pi^2}\sum_{k}  |y^N_{ik}|^2 \\
&\times C_{00}\left[m_Z^2,0,0, m^2_{H^\pm},m^2_{H^\pm}, m_{N_k}^2\right] ,\\
\label{eq:dGLeCT}
(\delta F_{L}^{Z\rightarrow l_il_i})^{ct} & = {\rm e}g^{Ze}_L \bigg[\frac{\delta_Z}{2}+\delta_{l_{Li}}+\delta{\rm e}+\frac{\delta_{AZ}}{2g^{Ze}_L}\\
&+\frac{1}{4s^3_Wc_W g^{Ze}_L}\left(\frac{\delta m^2_W}{m^2_W}-\frac{\delta m^2_Z}{m^2_Z}\right)\bigg].
\eea
Meanwhile, for SM neutrinos, we have
\bea
\label{eq:dGLnuCT}
 \delta F_{L}^{Z\rightarrow \nu_i \nu_i} & = \frac{{\rm e}g^{Z\nu}_L}{8 \pi^2}\sum_{k}  |y^N_{ik}|^2 \\
 &\times C_{00}\left[m_Z^2,0,0, m^2_{A^0},m^2_{H^0}, m_{N_k}^2\right],\\
 (\delta F_{L}^{Z\rightarrow \nu_i \nu_i})^{ct} & = {\rm e}g^{Z\nu}_{L}\bigg[\frac{\delta_Z}{2}+\delta_{\nu_{Li}} +\delta{\rm e}\\
 &+\frac{c^2_W-s^2_W}{2 s^2_W} \left(\frac{\delta m^2_W}{m^2_W}-\frac{\delta m^2_Z}{m^2_Z}\right)\bigg].
\eea
The full counterterm structures needed for $Z \rightarrow l^+ l^-$ and $Z \rightarrow \nu \nu$, in (\ref{eq:dGLeCT}) and (\ref{eq:dGLnuCT}), include the universal contribution from the IDM scalars in \ref{appIDMCT} (through wave-function renormalization of the $Z$) and a non-universal one provided in \ref{appCT}.
Analogous modifications affect the partial widths of the $W$-boson into leptons that is $W^{\pm}\rightarrow l^\pm \nu_l$:
\bea
\delta F_{L}^{W\rightarrow l_i\nu_i} & = \frac{{\rm e}g^{We\nu}_L }{8\pi^2}\sum_{k}  |y^N_{ik}|^2 \\
\times&\bigg( C_{00}\left[m_W^2,0,0, m^2_{H^\pm},m^2_{H^0}, m_{N_k}^2\right] \\&+C_{00}\left[m_W^2,0,0, m^2_{A^\pm},m^2_{H^\pm}, m_{N_k}^2\right] \bigg),\\
(\delta F_{L}^{W\rightarrow l_i\nu_i})^{ct} & = {\rm e}g^{We\nu}_L \bigg[\frac{\delta_W}{2}+\frac{\delta_{l_{Li}}}{2}+\frac{\delta_{\nu_{Li}}}{2}+\delta{\rm e}\\
&+\frac{c^2_W}{2s^2_W}\left(\frac{\delta m^2_W}{m^2_W}-\frac{\delta m^2_Z}{m^2_Z}\right)\bigg].\\
\text{Here $g^{We\nu}_L=-1/\sqrt{2}s_W$.}\hspace{-1cm}\nn
\eea

\bibliography{ScotoFCC}

\end{document}